\documentclass[aps,twocolumn,showpacs,preprintnumbers,amsmath,amssymb,
superscriptaddress,nofootinbib]{revtex4-1}

\usepackage{graphicx,epstopdf,amsmath,bm,amssymb,xcolor,xspace,braket}
\usepackage{hyperref}
\usepackage[ulem=normalem]{changes}

\begin{document}

\title{Neutron-proton pairing and double-beta decay in the interacting boson model}
\author{P.~Van~Isacker}
\affiliation{Grand Acc\'el\'erateur National d'Ions Lourds, CEA/DRF-CNRS/IN2P3,
Bvd Henri Becquerel, F-14076 Caen, France}

\author{J.~Engel}
\affiliation{Department of Physics and Astronomy, University of North Carolina, Chapel Hill, NC, 27516-3255, USA}

\author{K.~Nomura}
\affiliation{Grand Acc\'el\'erateur National d'Ions Lourds,
CEA/DRF-CNRS/IN2P3, Bvd Henri Becquerel, F-14076 Caen, France}
\affiliation{Physics Department, Faculty of Science, University of Zagreb, HR-10000 Zagreb, Croatia}
\affiliation{Center for Computational Sciences, University of Tsukuba, Tsukuba 305-8577, Japan}

\date{\today}  

\begin{abstract}
\noindent
{\bf Background:} The interacting boson model has been used extensively to
calculate the matrix elements governing neutrinoless double-beta decay.  Studies
within other models---the shell model, the quasiparticle random-phase
approximation, and nuclear energy-density functional theory---indicate that a
good description of neutron-proton pairing is essential for accurate
calculations of those matrix elements.  The usual interacting boson model is
based only on like-particle pairs, however, and the extent to which it captures
neutron-proton pairing is not clear.

\noindent
{\bf Purpose:} To determine whether neutron-proton pairing should be explicitly
included as neutron-proton bosons in interacting-boson-model calculations of
neutrinoless double-beta decay matrix elements.

\noindent
{\bf Method:} An isospin-invariant version of the nucleon-pair shell model is
applied to carry out shell-model calculations in a large space and in a
collective subspace, and to define effective operators in the latter.  A
democratic mapping is then used to define corresponding boson operators for the
interacting boson model, with and without an isoscalar neutron-proton pair
boson.  

\noindent
{\bf Results:} Interacting-boson-model calculations with and without the
isoscalar boson are carried out for nuclei near the beginning of the $pf$ shell,
with a realistic shell-model Hamiltonian and neutrinoless double-beta-decay
operator as the starting point.  Energy spectra and double-beta matrix elements
are compared to those obtained in the underlying shell model.

\noindent
{\bf Conclusions:} The isoscalar boson is not important for energy spectra but
improves the results for the double-beta matrix elements.  To be useful at the
level of precision we need, the mapping procedure must be further developed to
better determine the dependence of the boson Hamiltonian and decay operator on
particle number and isospin.  But the benefits provided by the isoscalar boson
suggest that through an appropriate combination of mappings and fitting, it
would make interacting-boson-model matrix elements more accurate for the heavier nuclei used in
experiments.
\end{abstract}

\maketitle

\section{Introduction}
\label{s_intro}

Experiments to measure the rate of neutrinoless double-beta ($0\nu\beta\beta$)
decay, in which two neutrons decay to two protons and two electrons, are growing
in number and expense~\cite{Henning16,DellOro16,Cremonesi14,GomezCadenas12}. A
nonzero rate would imply that neutrinos are Majorana particles~\cite{sch82} and
provide information about neutrino masses~\cite{bil87,avi08}, and possibly about
exotic new particles~\cite{Deppisch16}.  To efficiently plan and interpret the
experiments, however, one must know with reasonable accuracy the nuclear matrix
elements that govern the decay.  The matrix elements cannot be measured, and so
calculating them with controlled precision has become an important theoretical
goal. 

A variety of many-body methods have been loosed on the matrix element
problem~\cite{engel16}.  Of these, most are embedded in phenomenological models
that describe energy levels and other decay processes well in heavy nuclei.  The
interacting boson model (IBM)~\cite{Iachello87}, in which the fundamental
constituents are spin-zero and spin-two bosons that stand both for like-nucleon
pairs and the collective quadrupole degree of freedom, is a good
example~\cite{bar09,bar13,bar15}.  The version called
\mbox{IBM-2}~\cite{Arima77}, in which neutron and proton bosons are
distinguishable, successfully describes spectra and electromagnetic
transitions~\cite{Iachello87} plus single-$\beta$ decay~\cite{dellagiacoma89} in
a wide variety of nuclei.  

Because the IBM is equivalent to the Bohr-Mottelson collective model in certain
limits, one might expect the important physics in the IBM to mirror that in
approaches such as nuclear energy-density functional (EDF) theory, in which
quadrupole degrees of freedom are emphasized.  Similarly, because the IBM's
bosons are related to collective shell-model excitations, the important physics
in the IBM and the shell model should have something in common.  In both
EDF-based methods---in particular, the generator coordinate method (GCM)---and
in the shell model, $0\nu\beta\beta$ matrix elements tend to be too large unless
care is taken with the $J=1$, $T=0$ (isoscalar) pairing interaction. That
component of the nuclear Hamiltonian has been known to be important from QRPA
calculations~\cite{vog86,eng88} for nearly 30 years. More recently,
Ref.~\cite{hin14} showed that the inclusion of the isoscalar-pairing amplitude
as a generator coordinate reduces matrix elements significantly, and
Ref.~\cite{men16} showed that the isoscalar pairing in the Hamiltonian does the
same in the shell model.  Here we investigate whether it plays a similar role in
the IBM.  More explicitly, we ask whether we need to add a spin-1 isoscalar
boson to the model to avoid over-estimating $0\nu\beta\beta$ matrix elements,
and try to provide an answer.

To address the question, one would like to modify the IBM $0\nu\beta\beta$
calculations of Refs.~\cite{bar09,bar13,bar15} by adding the isoscalar spin-1
boson, which we will label $p$.
But because the IBM Hamiltonian is phenomenological, we do not know \textit{a
priori} how to add to and alter its Hamiltonian so as to correctly include the
physics of the new boson.  One might examine single-$\beta$ decay and other
processes in which the boson could play a significant role in order to pin down
new terms in the Hamiltonian, but that task is beyond what we are able to do
here.  Instead, we try to derive the boson Hamiltonian and the corresponding
decay operator, not in nuclei that actually undergo $0\nu\beta\beta$ decay, but
rather in light $pf$-shell nuclei for which exact shell model results are
available, both for spectra and for $0\nu\beta\beta$ matrix elements
\cite{men16}.  Since the IBM is supposed to represent collective dynamics in a
major shell, we construct a mapping of operators from the $pf$ shell to a set of
bosons, including not only the usual bosons of the \mbox{IBM-2} but
``neutron-proton'' bosons as well.  Ref.~\cite{bar09} developed a mapping to
obtain a boson $0\nu\beta\beta$ operator; here, starting from square one, we map
the Hamiltonian as well.  Though this procedure will not tell us how much a $p$
boson would change the realistic IBM $0\nu\beta\beta$ predictions, it will give
us a good idea of the extent to which a $p$ boson is required to faithfully
reflect shell-model results.

It is not obvious that the extra boson will improve the description of
$0\nu\beta\beta$ decay.  Certainly it will capture some of the isoscalar-pair
correlations that elude the \mbox{IBM-2}.  But because the product of two
isoscalar $J=1$ pair creation operators can be rewritten as a superposition of
products of proton pair creation operators and neutron pair creation operators,
some of the physics of isoscalar pairing is probably already in the
\mbox{IBM-2}.  Furthermore, the use of too many boson types can cause a model to
degrade for a related reason: an over-counting, roughly speaking, of the
independent collective degrees of freedom.  Thus, though an independent
isoscalar-pairing coordinate is clearly important in GCM calculations of
$0\nu\beta\beta$ matrix elements, an isoscalar boson may or may not improve the
corresponding IBM calculations. 

We carry out the mapping from shell model to IBM in two steps, the first of
which relies on the nucleon-pair shell model (NPSM)~\cite{Chen93,Chen97,Zhao14}
to define a collective subspace. We briefly review the NPSM in
Secs.~\ref{s_npsm} and~\ref{s_colsub}, with the main purpose of introducing
notation for an isospin-invariant formulation of the model~\cite{Fu13}; with a
collective subspace specified by the NPSM, we show in Sec.~\ref{s_effop} that we
can use the method of Suzuki and Lee~\cite{suzuki80} to construct effective
operators for the subspace.  The second step, which we describe in
Sec.~\ref{s_map}, involves the mapping of the (effective) Hamiltonian and the
$0\nu\beta\beta$-decay operator from the fermion to boson spaces.  In
Sec.~\ref{s_appli} we apply the formalism to the energy spectra of and
$0\nu\beta\beta$-transition strengths between nuclei in the lower part of the
$pf$ shell, taking into account correlations in the entire shell.  Finally, in
Sec.~\ref{s_conc}, we present our conclusions.

\section{The nucleon-pair shell model}
\label{s_npsm}

We introduce the following notation for pairs of fermions:
\begin{equation}
P^\dag_{\alpha\Gamma M_\Gamma}\equiv
(a^\dag_{\gamma_1}\times a^\dag_{\gamma_2})^{(\Gamma)}_{M_\Gamma}\,,
\label{e_pair}
\end{equation}
where $\gamma_i$ denotes the angular momentum $j_i$ and isospin $t_i$ (which is
always $\frac12$) of a single nucleon, $\Gamma$ stands for the coupled angular
momentum $J$ and isospin $T$ (which can be 0 or 1) of the nucleons with
$\gamma_1$ and $\gamma_2$ (with $\alpha$ standing for $\gamma_1,\gamma_2$), and
$M_\Gamma$ represents the corresponding projections (i.e.,
$M_\Gamma=M_JM_T$).  We write an arbitrary $n$-pair state of the NPSM as
\begin{widetext}
\begin{equation}
\ket{\alpha_1\Gamma_1\dots\alpha_n\Gamma_n\Lambda_2\dots \Lambda_n}\equiv
\left(\cdots\left(\left(P_{\alpha_1\Gamma_1}^\dag\times P_{\alpha_2\Gamma_2}^\dag\right)^{(\Lambda_2)}\times
P_{\alpha_3\Gamma_3}^\dag\right)^{(\Lambda_3)}\times\cdots\times
P_{\alpha_n\Gamma_n}^\dag\right)^{(\Lambda_n)}\ket{\rm O},
\label{e_bas1}
\end{equation}
\end{widetext}
or, 
\begin{equation}
\ket{P_r}\equiv\ket{\alpha_1\Gamma_1\dots\alpha_n\Gamma_n\Lambda_2\dots \Lambda_n},
\;r=1,\dots,\Theta\,,
\label{e_bas2}
\end{equation}
for short.  The state $\ket{\rm O}$ in Eq.~\eqref{e_bas1} is the bare fermion
vacuum, and the index $r$ in Eq.\ \eqref{e_bas2} stands for the set of quantum
numbers $\{\alpha_1\Gamma_1\dots\alpha_n\Gamma_n\Lambda_2\dots \Lambda_n\}$,
which specifies the character of the $n$ pairs $\alpha_q\Gamma_q$, the set
$\Lambda_q$ of intermediate angular momenta and isospins, and $\Lambda_n$, the
state's total angular momentum and isospin.

In general the basis in Eq.~\eqref{e_bas1} is non-orthonormal and overcomplete.
A calculation in this basis therefore requires the diagonalization of the
overlap matrix $\braket{P_r|P_s}$, the elements of which can be computed with a
recurrence relation presented in Ref.~\cite{Chen97} and generalized to include
isospin in Ref.~\cite{Fu13}.  Here we need matrix elements between one- and
two-pair states; these are trivial for $n=1$ and summarized in
Appendix~\ref{a_mes} for $n=2$.  Vanishing eigenvalues of the matrix
$\braket{P_r|P_s}$ indicate the overcompleteness of the pair basis. In a
subspace of $\Omega$ pair states in which all eigenvalues of the overlap matrix
are non-zero, one can construct an orthonormal set given by: 
\begin{equation}
\ket{\bar P_r}=
\sqrt{\frac{1}{O_r}}
\sum_{s=1}^\Omega C_{rs}\ket{P_s}\equiv
\sum_{s=1}^\Omega\bar C_{rs}\ket{P_s},
\;r=1,\dots,\Omega\,,
\label{e_obas}
\end{equation}
where $O_r$ is the $r^{\rm th}$ eigenvalue of the $\Omega\times\Omega$ overlap
matrix:
\begin{equation}
O_{rs}\equiv\braket{ P_r|P_s},
\;r,s=1,\dots,\Omega\,,
\label{e_over}
\end{equation}
and the coefficients $\{C_{rs},s=1,\dots,\Omega\}$ specify the corresponding
eigenvector.  If $\Omega$ is the dimension of the original shell-model space
$\mathbb H$, which we call the ``complete'' shell-model space, then the vectors
$\{\ket{\bar P_r},r=1,\dots,\Omega\}$ form a basis of ${\mathbb H}$. 

It is important to distinguish between the total number $\Theta$ of possible
$n$-pair states in Eq.~\eqref{e_bas2} and the number $\Omega\leq\Theta$ of
linearly independent states among them.  To apply the NPSM in a collective
subspace, one needs to expand an arbitrary $n$-pair state in terms of the
orthogonal basis states,
\begin{equation}
\ket{P_r}=
\sum_{s=1}^\Omega A_{rs}\ket{\bar P_s},
\;r=1,\dots,\Theta\,,
\label{e_expan1}
\end{equation}
where the coefficients $A_{rs}$ are given by 
\begin{equation}
A_{rs}\equiv\braket{ P_r|\bar P_s}=
\sum_{t=1}^\Omega\bar C_{st}\braket{ P_r|P_t},
\label{e_expan2}
\end{equation}
with $r=1,\dots,\Theta$ and $s=1,\dots,\Omega$. We assume here and henceforth
that the matrix elements are real, that is, that $\braket{
P_r|P_s}=\braket{P_s|P_r}$ and $\braket{\bar P_r|P_s}=\braket{P_s|\bar P_r}$.

An arbitrary shell-model operator $\hat T^{\rm f}$ (where f stands for
``fermion'') between two sets of orthogonal basis states,
\begin{equation}
\begin{split}
\ket{\bar P'_{r'}} &=
\sum_{s'=1}^{\Omega'}\bar C'_{r's'}\ket{P'_{s'}},
\;r'=1,\dots,\Omega'\,,\\
\ket{P''_{r''}} &=
\sum_{s''=1}^{\Omega''}\bar C''_{r''s''}\ket{P''_{s''}},
\;r''=1,\dots,\Omega''\,,
\end{split}
\label{e_orth-nonorth}
\end{equation}
has the matrix elements
\begin{equation}
\begin{split}
\braket{\bar P'_{r'}|\hat T^{\rm f}|\bar P''_{r''}} &=
\sum_{s'=1}^{\Omega'}\sum_{s''=1}^{\Omega''}\bar C'_{r's'}\bar C''_{r''s''}
\braket{P'_{s'}|\hat T^{\rm f}|P''_{s''}}\\ &=
(\bar{\pmb{C}}'\times\pmb{T}^{\rm f}\times\bar{\pmb{C}}''^T)_{r'r''}\,.
\end{split}
\label{e_tmat}
\end{equation}
Here $\pmb{M}^T$ is the transposed matrix of $\pmb{M}$ and $\bar{\pmb{C}}'$,
$\pmb{T}^{\rm f}$, and $\bar{\pmb{C}}''$ are the following matrices:
\begin{equation}
\begin{split}
\bar{\pmb{C}}' &:
\{\bar C'_{r's'},\;r'=1,\dots,\Omega',\;s'=1,\dots,\Omega'\}\,,\\
\pmb{T}^{\rm f} &:
\{\braket{ P'_{s'}|\hat T^{\rm f}|P''_{s''}},
\;s'=1,\dots,\Omega',\;s''=1,\dots,\Omega''\}\,,\\
\bar{\pmb{C}}'' &:
\{\bar C''_{r''s''},\;r''=1,\dots,\Omega'',\;s''=1,\dots,\Omega''\}\,.
\end{split}
\label{e_matrix1}
\end{equation}

For the Hamiltonian operator, $\hat T^{\rm f}=\hat H^{\rm f}$, the dimensions in
bra and ket of Eq.~(\ref{e_tmat}) are the same, $\Omega'=\Omega''\equiv\Omega$,
and the diagonalization of the $\Omega\times\Omega$ Hamiltonian matrix leads to
the eigenstates
\begin{equation}
\ket{\bar E_t}=
\sum_{r=1}^\Omega E_{tr}\ket{\bar P_r},
\;t=1,\dots,\Omega\,,
\label{e_vec}
\end{equation}
where the coefficients $\{E_{tr},r=1,\dots,\Omega\}$ are the components of the
eigenvector associated with the eigenvalue $E_t$. If $\hat T^{\rm f}$ is another
operator, e.g.\ a transition operator, its action on eigenstates of $\hat H^{\rm
f}$ in the complete Hilbert space ${\mathbb H}$ is given by
\begin{equation}
\begin{split}
\braket{\bar E'_{t'}|\hat T^{\rm f}|\bar E''_{t''}} &=
\sum_{r'=1}^{\Omega'}\sum_{r''=1}^{\Omega''}E'_{t'r'}E''_{t''r''}
\braket{\bar P'_{r'}|\hat T^{\rm f}|\bar P''_{r''}}\\ &=
(\pmb{E}'\times\bar{\pmb{C}}'\times\pmb{T}^{\rm f}\times\bar{\pmb{C}}''^T\times\pmb{E}''^T)_{t't''}\,,
\end{split}
\label{e_tmat3}
\end{equation}
with $t'=1,\dots,\Omega'$ and $t''=1,\dots,\Omega''$.

All this shows us that it is possible to carry out standard shell-model
calculations in the NPSM, albeit in a complicated way.  The advantage of the
NPSM is that it allows a truncation to a shell-model subspace constructed in
terms of collective pairs.

\section{Collective subspace}
\label{s_colsub}
A collective fermion pair is a superposition of pairs built from orbits with
different $\gamma_1$ and $\gamma_2$, all coupled to the same $\Gamma$.  It can
be specified by its coefficients $\alpha^\Gamma_{\gamma_1\gamma_2}$:
\begin{equation}
B^\dag_{\alpha\Gamma M_\Gamma}\equiv
\sum_{\gamma_1\gamma_2}
\alpha^\Gamma_{\gamma_1\gamma_2}(a^\dag_{\gamma_1}\times
a^\dag_{\gamma_2})^{(\Gamma)}_{M_\Gamma}\,,
\label{e_cpair}
\end{equation}
where the subscript $\alpha$ is to emphasize the dependence of the pair on its
coefficients.  After selecting a particular set of collective pairs
$\{B_\Gamma\}$, we can construct NPSM states from them, {\it viz.}
\begin{widetext}
\begin{equation}
\ket{\alpha_1\Gamma_1\dots\alpha_n\Gamma_n\Lambda_2\dots \Lambda_n}\equiv
\left(\cdots\left(\left(B_{\alpha_1\Gamma_1}^\dag\times B_{\alpha_2\Gamma_2}^\dag\right)^{(\Lambda_2)}\times
B_{\alpha_3\Gamma_3}^\dag\right)^{(\Lambda_3)}\times\cdots\times
B_{\alpha_n\Gamma_n}^\dag\right)^{(\Lambda_n)}\ket{{\rm O}}.
\label{e_fbas1}
\end{equation}
\end{widetext}
Although the formalism does not require it, we shall henceforth consider only one
collective pair for a given $\Gamma$ (i.e., $J$ and $T$).  This step
enables us to dispense with the indices $\alpha_q$, reducing the necessary
labels to $\{\Gamma_1\dots\Gamma_n\Lambda_2\dots \Lambda_n\}$ and leading to the
abbreviation
\begin{equation}
\ket{B_i}\equiv\ket{\Gamma_1\dots\Gamma_n\Lambda_2\dots \Lambda_n},
\;i=1,\dots,\omega\,,
\label{e_fbas2}
\end{equation}
where $\omega$ is the number of couplings $\{\Gamma_1\dots\Gamma_n\Lambda_2\dots \Lambda_n\}$.

The collective NPSM states in Eq.~\eqref{e_fbas1} can be expressed as linear
combinations of the non-collective ones in Eq.~\eqref{e_bas1}:
\begin{equation}
\begin{split}
\ket{B_i} &=
\sum_{r=1}^\Theta a_{ir}\ket{P_r}=
\sum_{r=1}^\Theta\sum_{s=1}^\Omega a_{ir}A_{rs}\ket{\bar P_s}\\ &=
\sum_{s=1}^\Omega(\pmb{a}\times\pmb{A})_{is}\ket{\bar P_s},
\;i=1,\dots,\omega\,,
\end{split}
\label{e_fbas3}
\end{equation}
with coefficients $a_{ir}$ that are functions of
$\alpha^\Gamma_{\gamma_1\gamma_2}$.  Results analogous to those in Sec.\
\ref{s_npsm} can now be obtained by diagonalizing the collective overlap matrix
\begin{equation}
\begin{split}
o_{ij} &\equiv
\braket{ B_i|B_j}=
\sum_{r=1}^\Omega(\pmb{a}\times\pmb{A})_{ir}(\pmb{a}\times\pmb{A})_{jr}\\ &=
(\pmb{a}\times\pmb{A}\times\pmb{A}^T\times\pmb{a}^T)_{ij},
\;i,j=1,\dots,\omega\,.
\end{split}
\label{e_fover}
\end{equation}
The number of linearly independent vectors $\ket{B_i}$ is given by the number of
non-zero eigenvalues of the matrix $o_{ij}$.
We assume here that all eigenvalues of the matrix~(\ref{e_fover}) are non-zero, which will
be the case for any reasonable choice of the collective subspace.  As we have
noted, the NPSM is interesting because when one restricts oneself to a set of
collective pairs, the resulting space has a much lower dimension than does
$\mathbb H$ itself, \textit{i.e.}, $\omega\ll\Omega$.

To carry out a shell-model calculation in the collective subspace, we employ
notation that is analogous to what we used for the full space $\mathbb H$.
Thus, we work with orthonormal states 
\begin{equation}
\begin{split}
\ket{\bar B_i} &=
\sqrt{\frac{1}{o_i}}
\sum_{j=1}^\omega c_{ij} \ket{B_j}\equiv
\sum_{j=1}^\omega\bar c_{ij}\ket{B_j}\\ &=
\sum_{r=1}^\Omega(\bar{\pmb{c}} \times \pmb{a}\times\pmb{A})_{ir} 
\ket{\bar P_r} \equiv \sum_{r=1}^\Omega b_{ir}\ket{\bar P_r},
\end{split}
\label{e_fobas}
\end{equation}
where $o_i$ is the $i^{\rm th}$ eigenvalue of the overlap matrix
($i=1,\dots,\omega$) and the coefficients $\{c_{ij}, j=1,\dots,\omega\}$ make
up the corresponding eigenvector.  A shell-model operator $\hat T^{\rm f}$ has
the matrix elements
\begin{equation}
\begin{split}
\braket{\bar B'_{i'} | \hat T^{\rm f}|\bar B''_{i''}} &=
\sum_{r'=1}^{\Omega'}\sum_{r''=1}^{\Omega''}b_{i'r'}b_{i''r''}
\braket{\bar P'_{r'}|\hat T^{\rm f}|\bar P''_{r''}}\\ &=
(\pmb{b}'\times\bar{\pmb{C}}'\times\pmb{T}^{\rm f}\times
\bar{\pmb{C}}''^T\times\pmb{b}''^T)_{i'i''}\,,
\end{split}
\label{e_ftmat}
\end{equation}
with $i'=1,\dots,\omega'$ and $i''=1,\dots, \omega''$.  For the
Hamiltonian operator, $\hat T^{\rm f}=\hat H^{\rm f}$, the
matrix~(\ref{e_ftmat}) has dimension $\omega\times\omega$ and its
diagonalization leads to the eigenstates
\begin{equation}
\ket{\bar e_k}=
\sum_{i=1}^\omega e_{ki}\ket{\bar B_i}=
\sum_{r=1}^\Omega (\pmb{e}\times\pmb{b})_{kr} \ket{\bar P_r},
\label{e_fvec}
\end{equation}
with $k=1,\dots,\omega$ and with the coefficients
$\{e_{ki},i=1,\dots,\omega\}$ given by the components of the eigenvector
associated with the eigenvalue $e_k$.  If $\hat T^{\rm f}$ is some other
operator, its matrix elements in the basis of eigenstates of $\hat H^{\rm f}$
are given by
\begin{equation}
\braket{\bar e'_{k'}|\hat T^{\rm f}|\bar e''_{k''}}=
(\pmb{e}'\times\pmb{b}'\times\bar{\pmb{C}}'\times\pmb{T}^{\rm f}
\times\bar{\pmb{C}}''^T\times\pmb{b}''^T\times\pmb{e}''^T)_{k'k''}\,,
\label{e_ftmat3}
\end{equation}
with $k'=1,\dots,\omega'$ and $k''=1,\dots,\omega''$.

\section{Effective shell-model operators in a collective subspace}
\label{s_effop}
The eigenspectrum of $\hat H^{\rm f}$ and the matrix elements of transition
operators $\hat T^{\rm f}$ in the restricted Hilbert space differ from the
corresponding eigenspectrum and matrix elements of the operators in the complete
Hilbert space ${\mathbb H}$.  We need an effective Hamiltonian $\hat H^{\rm f}_{\rm eff}$
and, more generally, effective operators $\hat T^{\rm f}_{\rm eff}$ in the restricted
Hilbert space that preserve the original eigenvalues and
matrix elements.  We begin their construction by letting ${\mathbb H}_P$ be the
restricted Hilbert space and ${\mathbb H}_Q$ the excluded Hilbert space, with
${\mathbb H}={\mathbb H}_P\cup{\mathbb H}_Q$.  The operators $\hat P$ and $\hat Q$
project onto the corresponding spaces, so that $\hat P{\mathbb H}={\mathbb H}_P$
and $\hat Q{\mathbb H}={\mathbb H}_Q$.  The eigenstates of $\hat H^{\rm f}$
in ${\mathbb H}$ are given by $\{\ket{\bar E_t},t=1,\dots,\Omega\}$ and
those of $\hat H^{\rm f}$ in ${\mathbb H}_P$ by $\{\ket{\bar e_k},
k=1,\dots,\omega\}$.  For each eigenstate $\ket{\bar e_k}$ we identify a
corresponding eigenstate $\ket{\bar E_k}$, usually by requiring a maximum
overlap
\begin{equation}
\braket{\bar E_k|\bar e_k}=
\sum_{r=1}^\Omega E_{tr}(\pmb{e}\times\pmb{b})_{kr}=
(\pmb{E}\times\pmb{b}^T\times\pmb{e}^T)_{tk}\,.
\label{e_eeover}
\end{equation}
This procedure defines a set of $\omega$ eigenstates $\{\ket{\bar
E_{t_k}},k=1,\dots,\omega\}$ and an associated $\omega\times\Omega$ matrix
$\tilde{\pmb{E}}$ with the elements
\begin{equation}
\tilde{\pmb{E}}:
\{\tilde E_{kr}\equiv E_{t_kr},\;k=1,\dots,\omega,\;r=1,\dots,\Omega\}\,.
\label{e_emat}
\end{equation}
For each of the eigenstates $\ket{\bar E_{t_k}}$ we define its component in ${\mathbb H}_P$,
\begin{equation}
\ket{e_k}\equiv\hat P\ket{\bar E_{t_k}},
\;k=1,\dots,\omega\,,
\label{e_pcomp}
\end{equation}
and assume that the states $\{\ket{e_k},k=1,\dots,\omega\}$ are linearly
independent and therefore span the entire restricted Hilbert space ${\mathbb H}_P$.

We use the method of Suzuki and Lee~\cite{suzuki80} to determine effective
operators in ${\mathbb H}_P$.  The method employs an operator $\hat\eta$ that
maps states in ${\mathbb H}_P$ to states in ${\mathbb H}_Q$ such that
\begin{equation}
\hat\eta\ket{e_k}=\hat Q\ket{\bar E_{t_k}},
\;k=1,\,\dots,\,\omega\,.
\label{e_eta}
\end{equation}
Since $\hat\eta=\hat Q\hat\eta\hat P$, the operator $\hat\eta$ satisfies the
relations
\begin{equation}
\hat P\hat\eta=\hat\eta\hat Q=0\,,\quad
\hat\eta\hat P=\hat Q\hat\eta=\hat\eta\,,
\quad \hat\eta^2=0\,.
\label{e_eta1}
\end{equation}
It follows that
\begin{equation}
\ket{\bar E_{t_k}}=
(\hat P+\hat Q)\ket{\bar E_{t_k}}=
\ket{e_k}+\hat\eta\ket{e_k}=
(\hat I+\hat\eta)\ket{e_k},
\label{e_eta2}
\end{equation}
and, inversely, that
\begin{equation}
\ket{e_k}=
(\hat I-\hat\eta)(\hat I+\hat\eta)\ket{e_k}=
(\hat I-\hat\eta)\ket{\bar E_{t_k}}.
\label{e_eta3}
\end{equation}
The transformed (non-hermitian) Hamiltonian $\hat{\cal H}^{\rm f}\equiv(\hat
I-\hat\eta)\hat H^{\rm f}(\hat I+\hat\eta)$ satisfies the relation 
\begin{equation}
\hat{\cal H}^{\rm f}\ket{e_k}=
(\hat I-\hat\eta)\hat H^{\rm f}\ket{\bar E_{t_k}}=
E_{t_k}(\hat I-\hat\eta)\ket{\bar E_{t_k}}=
E_{t_k}\ket{e_k},
\label{e_hamtf}
\end{equation}
which shows that the states $\{\ket{e_k},k=1,\dots,\omega\}$ are eigenstates
of $\hat{\cal H}^{\rm f}$, with eigenvalues $\{E_{t_k},k=1,\dots,\omega\}$.

A matrix element of $\hat\eta$ is nonzero only if the bra is in ${\mathbb H}_Q$ and
the ket in ${\mathbb H}_P$.  The operator is therefore determined by the matrix
elements
\begin{equation}
\braket{\bar P_r|\hat\eta|\bar B_i},
\;r=1,\dots,\Omega\,,
\;i=1,\dots,\omega\,,
\label{e_mateta1}
\end{equation}
where $\ket{\bar P_r}\in{\mathbb H}$ and $\ket{\bar B_i}\in{\mathbb H}_P$.
To calculate the matrix elements in Eq.~\eqref{e_mateta1}, we first note that
although the states $\{\ket{e_k},k=1,\dots,\omega\}$ span the entire Hilbert
space ${\mathbb H}_P$, they do not form an orthonormal basis.  A bi-orthogonal
basis $\{\bra{\tilde e_k},k=1,\dots,\omega\}$ can be defined such that
\begin{equation}
\braket{\tilde e_k|e_{k'}}=\delta_{kk'}\,,
\label{e_biortho}
\end{equation}
which implies that the operator $\hat P$, which is nothing but the identity operator on ${\mathbb H}_P$,
can be written as
\begin{equation}
\hat P=\sum_{k=1}^\omega\ket{e_k}\bra{\tilde e_k}.
\label{e_ident}
\end{equation}
Thus we have
\begin{equation}
\braket{\bar P_r|\hat\eta|\bar B_i}=
\sum_{k=1}^\omega\braket{\bar P_r|\hat\eta|e_k}\braket{\tilde e_k|\bar B_i}.
\label{e_mateta2}
\end{equation}
The first matrix element in this sum can be written as
\begin{equation}
\braket{\bar P_r|\hat\eta|e_k}=
\braket{\bar P_r|\bar E_{t_k}}-
\sum_{i=1}^\omega\braket{\bar P_r|\bar B_i}\braket{\bar B_i|\bar E_{t_k}},
\label{e_mateta4}
\end{equation}
where we have used the relation $\hat P=\sum\ket{\bar B_i}\bra{\bar B_i}$.
With the help of Eqs.~\eqref{e_vec} and \eqref{e_fobas} we deduce the relations 
\begin{equation}
\braket{\bar P_r|\bar E_{t_k}}=\tilde E_{kr}\,,
\;\braket{\bar P_r|\bar B_i}=b_{ir}\,,
\;\braket{\bar B_i|\bar E_{t_k}}=
(\pmb{b}\times\tilde{\pmb{E}}^T)_{ik}\,,
\label{e_mateta5}
\end{equation}
which lead to the following expression for the first matrix element in the sum
in Eq.~\eqref{e_mateta2}:
\begin{equation}
\braket{\bar P_r|\hat\eta|e_k}=
\tilde E_{kr}-
(\tilde{\pmb{E}}\times\pmb{b}^T\times\pmb{b})_{kr}\,.
\label{e_mateta6}
\end{equation}
To determine the second matrix element in the sum in Eq.~\eqref{e_mateta2}, we
note that
\begin{equation}
\braket{\bar B_i|\bar B_j}=
\sum_{k=1}^\omega\braket{\bar B_i|e_k}\braket{\tilde e_k|\bar B_j}=
\delta_{ij}\,,
\label{e_mateta7}
\end{equation}
which implies that the matrix $\braket{\tilde e_k|\bar B_i}$ is the inverse
of the matrix with the elements
\begin{equation}
d_{ik}\equiv
\braket{\bar B_i|e_k}=
\braket{\bar B_i|\hat P|\bar E_{t_k}}=
(\pmb{b}\times\tilde{\pmb{E}}^T)_{ik}\,.
\label{e_mateta8}
\end{equation}
We conclude that the matrix elements of the operator $\hat\eta$ are
\begin{equation}
\eta_{ri}\equiv
\braket{\bar P_r|\hat\eta|\bar B_i}=
\left((\pmb{I}_\Omega-\pmb{b}^T\times\pmb{b})\times
\tilde{\pmb{E}}^T\times\pmb{d}^{-1}\right)_{ri}\,,
\label{e_mateta9}
\end{equation}
with $r=1,\dots,\Omega$ and $i=1,\dots,\omega$, and $\pmb{I}_\Omega$ given by the
$\Omega\times\Omega$ identity matrix.  The matrix elements $\eta_{ri}$ can be
defined entirely in terms of the $\Omega^2$ NPSM overlap matrix elements in
Eq.~\eqref{e_over}.

With an expression for the matrix elements of $\hat{\eta}$, we are now finally
in a position to define effective operators for the collective subspace.  To any
operator $\hat T^{\rm f}$, which acts on states in a Hilbert space ${\mathbb H}''$
to give states in a Hilbert space ${\mathbb H}'$, there corresponds an
effective, hermitian operator~\cite{Navratil93}
\begin{equation}
\hat T^{\rm f}_{\rm eff}=
\hat P'\hat T'^{-1/2}_\eta
(\hat I'+\hat\eta'^\dag)\hat T^{\rm f}(\hat I''+\hat\eta'')
\hat T''^{-1/2}_\eta\hat P''\,,
\label{e_teff1}
\end{equation}
where
\begin{equation}
\hat T'_\eta\equiv(\hat I' +\hat\eta'^\dag\hat\eta')\,,\quad
\hat T''_\eta\equiv(\hat I'' +\hat\eta''^\dag\hat\eta'')\,.
\label{e_teff2}
\end{equation}
The operator $\hat T^{\rm f}_{\rm eff}$ acts on states in the restricted Hilbert
space ${\mathbb H}''_P$ to give states in the restricted Hilbert space ${\mathbb H}'_P$. 
Its matrix elements involve sums over the inverse square root of the
matrix
\begin{equation}
\braket{\bar B_i|\hat T_\eta|\bar B_j}=
(\pmb{I}_\omega+\pmb{\eta}^T\times\pmb{\eta})_{ij}\,,
\;i,j=1,\dots,\omega\,. 
\label{e_teff3}
\end{equation}
Since $\hat\eta$ defines a positive-definite metric, the square root of the
matrix in Eq.\ \eqref{e_teff3} can be taken through diagonalization.  And the
matrix elements of the other part of the operator in Eq.\ \eqref{e_teff1} are
given by
\begin{equation}
\begin{split}
&\braket{\bar B'_{j'}|(\hat I'+\hat\eta'^\dag)\hat T^{\rm f}(\hat I''+\hat\eta'')|\bar B''_{j''}}\\
&=((\pmb{b}'+\pmb{\eta}'^T)\times\bar{\pmb{C}}'\times\pmb{T}^{\rm
f}\times\bar{\pmb{C}}''^T\times(\pmb{b}''^T+\pmb{\eta}''))_{j'j''}\,,
\end{split}
\label{e_teff4}
\end{equation}
with $j'=1,\dots,\omega'$ and $j''=1,\dots,\omega''$.

Equation~\eqref{e_teff1} gives the effective version of any operator.  For the
Hamiltonian operator an alternative expressions exists~\cite{Navratil93}:
\begin{equation}
\hat H^{\rm f}_{\rm eff}\equiv
\hat P'\hat T'^{+1/2}_\eta
\hat H^{\rm f}(\hat I''+\hat\eta'')
\hat T''^{+1/2}_\eta\hat P''\,,
\label{e_heff1}
\end{equation}
which involves the matrix elements
\begin{equation}
\begin{split}
&\braket{\bar B'_{j'}|\hat H^{\rm f}(\hat I''+\hat\eta'')|\bar B''_{j''}}\\
&=(\pmb{b}'\times\bar{\pmb{C}}'\times\pmb{H}^{\rm
f}\times\bar{\pmb{C}}''^T\times(\pmb{b}''^T+\pmb{\eta}''))_{j'j''}\,,
\end{split}
\label{e_heff3}
\end{equation}
with $j'=1,\dots,\omega'$ and $j''=1,\dots,\omega''$.

\section{Mapping to boson operators}
\label{s_map}
To study the IBM, we need to map the collective fermion pairs onto bosons.
Given a choice of collective pairs $\{B_\Gamma\}$ we can introduce a
corresponding set of bosons $\{b_\Gamma\}$.  The boson analog of the state in
Eq.\ \eqref{e_fbas1} is
\begin{widetext}
\begin{equation}
\ket{b_i}\equiv
\left(\cdots\left(\left(b_{\Gamma_1}^\dag\times b_{\Gamma_2}^\dag\right)^{(\Lambda_2)}\times
b_{\Gamma_3}^\dag\right)^{(\Lambda_3)}\times\cdots\times
b_{\Gamma_n}^\dag\right)^{(\Lambda_n)}\ket{\rm o},
\;i=1,\dots,\omega\,,
\label{e_bbas}
\end{equation}
\end{widetext}
where the $\ket{\rm o}$ is the boson vacuum and the index $i$ is again short-hand
for the labels $\{\Gamma_1\dots\Gamma_n\Lambda_2\dots \Lambda_n\}$.  In general,
the boson states in Eq.~\eqref{e_bbas} do not form an orthonormal basis, leading
to complications described in Ref.~\cite{Isacker13}.  For a mapping that is
limited to operators with at most two-body terms between the bosons, however,
only boson states with $n=1$ and $n=2$ are needed, in which case the overlap
matrix $\braket{b_i|b_j}$ is diagonal and no such complications arise.  The
boson states that correspond to the orthogonalized fermion-pair states are
obtained from the unitary transformation
\begin{equation}
\ket{\bar b_i}=
\sum_{j=1}^\omega c_{ij}\ket{b_j},
\;i=1,\dots,\omega\,,
\label{e_bobas}
\end{equation}
with the coefficients $c_{ij}$ taken from the orthogonalization process in the
collective fermion subspace [see Eq.~\eqref{e_fobas}].

In the standard method for carrying out the boson mapping, introduced in the IBM
by Otsuka, Arima, and Iachello (OAI)~\cite{Otsuka78}, one uses the Gram-Schmidt
procedure to orthogonalize the fermion basis (not the boson basis).  The
procedure proposed in Eq.~(\ref{e_bobas}), known as democratic
mapping~\cite{Skouras90}, is different from OAI because it relies on the
diagonalization of the overlap matrix for the non-orthogonal fermion basis.  It
is thus similar to L\"owdin's symmetric orthogonalization
procedure~\cite{Lowdin50}, introduced in quantum chemistry, which yields the
orthogonal basis that is ``closest'' to the original non-orthogonal
one~\cite{Carlson57,Mayer02}.  The democratic mapping has the additional
advantage that it imposes no Graham-Schmidt-like hierarchy on the basis states,
and as a result is more useful for models with several kinds of bosons, such as
those to be considered here, for which an ordering would be arbitrary.  

In the democratic mapping the boson image $\hat T^{\rm b}$ of a fermion
operator, which can be $\hat T^{\rm f}$ or its effective version $\hat T^{\rm f}_{\rm eff}$,
is determined by the relation 
\begin{equation}
\braket{\bar b'_{i'}|\hat T^{\rm b}|\bar b''_{i''}}\doteq
\braket{\bar B'_{i'}|\hat T^{\rm f}|\bar B''_{i''}},
\;i^{\prime,\prime\prime}=1,\dots,\omega^{\prime,\prime\prime}\,,
\label{e_map1}
\end{equation}
where the symbol $\doteq$ indicates that the equality holds by virtue of the
mapping.  In terms of the original basis states $\ket{b_i}$, this relation leads
to the boson matrix elements
\begin{equation}
\begin{split}
&\braket{ b'_{i'}|\hat T^{\rm b}|b''_{i''}}\doteq
\sum_{j'=1}^{\omega'}\sum_{j''=1}^{\omega''}c'_{j'i'}c''_{j''i''}
\braket{\bar B'_{j'}|\hat T^{\rm f}|\bar B''_{j''}}\\
&=(\pmb{c}'^T\times\pmb{b}'\times\bar{\pmb{C}}'\times\pmb{T}^{\rm
f}\times\bar{\pmb{C}}''^T\times\pmb{b}''^T\times\pmb{c}'')_{i'i''}\,,
\end{split}
\label{e_map2}
\end{equation}
with $i'=1,\dots,\omega'$ and $i''=1,\dots,\omega''$.

In the most general mapping several technical issues arise, including the
elimination of spurious boson states and ambiguities for $n>2$, which are
described in Ref.~\cite{Isacker13}.  No such difficulties exist for a mapping to
a collective subspace with no more than two-body operators, $n\leq2$.  Even so,
one must carefully define boson operators order by order.  For example, for the
Hamiltonian operator, $\hat T=\hat H$, one determines the single-boson energy
from
\begin{equation}
\epsilon_\Gamma\equiv
\braket{ b_\Gamma|\hat H^{\rm b}|b_\Gamma}=
\braket{\bar b_\Gamma|\hat H^{\rm b}|\bar b_\Gamma}\doteq
\braket{\bar B_\Gamma|\hat H^{\rm f}|\bar B_\Gamma},
\label{e_hme1}
\end{equation}
and subsequently the two-body part of the boson Hamiltonian from
\begin{equation}
\begin{split}
&\upsilon^{\Lambda}_{\Gamma'_1\Gamma'_2\Gamma''_1\Gamma''_2} \equiv
\braket{b_{\Gamma'_1}b_{\Gamma'_2};\Lambda|\hat H^{\rm b}_2|
b_{\Gamma''_1}b_{\Gamma''_2};\Lambda}\\
&=
\braket{b_{\Gamma'_1}b_{\Gamma'_2};\Lambda|\hat H^{\rm b}|b_{\Gamma''_1}b_{\Gamma''_2};\Lambda}
-(\epsilon_{\Gamma'_1} + \epsilon_{\Gamma'_2})
\delta_{\Gamma'_1\Gamma''_1}\delta_{\Gamma'_2\Gamma''_2}\,,
\end{split}
\label{e_hme2}
\end{equation}
where we have assumed that the two-boson states are normalized and that the
labels $\Gamma'_i$ and $\Gamma''_i$ appear in some standard order (i.e.,
$\Gamma'_1\leq\Gamma'_2$ and $\Gamma''_1\leq\Gamma''_2$).  Similar reasoning
leads to somewhat more complicated expressions for the $0\nu\beta\beta$-decay operator
because it is non-scalar in isospin.  The order-by-order mapping of non-scalar
operators is explained in Appendix~\ref{a_order}.

\section{Application to $0\nu\beta\beta$ decay in the $pf$ shell}
\label{s_appli}
To test the role of an isoscalar-pair boson in the IBM, we now apply the
formalism presented above to nuclei in the lower part of the $pf$ shell.  We
will map both the Hamiltonian and the $0\nu\beta\beta$ operator from the $pf$
shell to a system of bosons. 

\begin{table*}[t]
\centering
\caption{Two-boson interaction matrix elements $\braket{ JT|\hat H^{\rm b}_2|JT}$
(in MeV), mapped from the KB3G interaction (IBMb, top) and from
the effective fermion interaction (IBMe, bottom).}
\label{t_v3}
\begin{ruledtabular}
\begin{tabular}{rcccccccccc}
&&\multicolumn{3}{c}{$J=0$}&$J=1$&\multicolumn{3}{c}{$J=2$}&$J=3$&$J=4$\\
\cline{3-5}\cline{7-9} $T$&~~~&
$\braket{ s^2|\hat V|s^2}$&
$\braket{ d^2|\hat V|d^2}$&
$\braket{ s^2|\hat V|d^2}$&
$\braket{ d^2|\hat V|d^2}$&
$\braket{ sd|\hat V|sd}$&
$\braket{ d^2|\hat V|d^2}$&
$\braket{ sd|\hat V|d^2}$&
$\braket{ d^2|\hat V|d^2}$&
$\braket{ d^2|\hat V|d^2}$\\
\hline
$0$&&$\begin{array}{c}-4.440\\[-3pt]-\mbox{\sl 6.379}\end{array}$&
$\begin{array}{c}-4.258\\[-3pt]-\mbox{\sl 5.013}\end{array}$&
$\begin{array}{c}-3.402\\[-3pt]-\mbox{\sl 3.489}\end{array}$&---&
$\begin{array}{c}-6.350\\[-3pt]-\mbox{\sl 8.263}\end{array}$&
$\begin{array}{c}-4.916\\[-3pt]-\mbox{\sl 7.873}\end{array}$&
$\begin{array}{c}\phantom{-}0.661\\[-3pt]-\mbox{\sl 0.563}\end{array}$&---&
$\begin{array}{c}-5.116\\[-3pt]-\mbox{\sl 8.642}\end{array}$\\
$1$&&---&---&---&
$\begin{array}{c}-3.029\\[-3pt]-\mbox{\sl 4.861}\end{array}$&
$\begin{array}{c}-2.284\\[-3pt]-\mbox{\sl 4.224}\end{array}$&---&---&
$\begin{array}{c}-3.106\\[-3pt]-\mbox{\sl 4.831}\end{array}$&---\\
$2$&&$\begin{array}{c}\phantom{-}2.985\\[-3pt]\phantom{-}\mbox{\sl 1.438}\end{array}$&
$\begin{array}{c}\phantom{-}3.229\\[-3pt]\phantom{-}\mbox{\sl 1.699}\end{array}$&
$\begin{array}{c}\phantom{-}3.490\\[-3pt]\phantom{-}\mbox{\sl 2.072}\end{array}$&---&
$\begin{array}{c}\phantom{-}0.896\\[-3pt]\phantom{-}\mbox{\sl 0.473}\end{array}$&
$\begin{array}{c}\phantom{-}0.899\\[-3pt]\phantom{-}\mbox{\sl 0.297}\end{array}$&
$\begin{array}{c}-0.822\\[-3pt]-\mbox{\sl 0.520}\end{array}$&---&
$\begin{array}{c}\phantom{-}0.719\\[-3pt]-\mbox{\sl 0.282}\end{array}$\\
\end{tabular}
\end{ruledtabular}
\end{table*}

Like Ref.~\cite{men16}, we use two fermion Hamiltonians:  the modified Kuo-Brown
Hamiltonian KB3G~\cite{Poves01}, which is used throughout the $pf$
shell~\cite{cau05}, and a multi-separable collective approximation to
it~\cite{duf96}.  The collective Hamiltonian contains a monopole term, isovector
pairing, isoscalar paring, a quadrupole-quadrupole interaction, and a
particle-hole spin-isospin interaction.  Its virtue is that it allows us to
selectively remove particular pieces of the Hamiltonian, for example isoscalar
pairing, which has been shown to be important~\cite{men16} and which we are
interested in investigating within the IBM.  The detailed results we report now,
however, are produced by KB3G. We will turn to the collective interaction
towards the end of this section.

We now construct two boson models designed to test the ability of the IBM, with
and without an explicit isoscalar-pair degree of freedom, to capture the physics
important to $0\nu\beta\beta$ decay.  The first consists of isovector bosons
with angular momentum $\ell=0$ ($s$) and $\ell=2$ ($d$), and with isospin $t=1$.
This set corresponds to the isospin-invariant version of the IBM known as
\mbox{IBM-3}~\cite{elliott80}.  In the second model the set is enlarged by
adding an isoscalar ($p$) boson with $\ell=1$ and $t=0$ (and positive parity).
The result is not the full SU(4)-invariant version of the interacting boson
model known as \mbox{IBM-4}~\cite{elliott81}, but it incorporates that model's
most important isoscalar correlations and therefore is situated somewhere
between \mbox{IBM-3} and \mbox{IBM-4}.  We refer generically to the boson models
here simply as the IBM.  To refer specifically to the versions without or with
the isoscalar boson, we use the terms IBM and \mbox{$p$-IBM}.

\subsection{Mapping of the Hamiltonian}
\label{ss_maph}
We use the order-by-order mapping described earlier to obtain the boson
Hamiltonian from two- and four-nucleon systems, that is, from the $A=42$ and
$A=44$ nuclei.  The first yields the boson energies, which turn out to be
$\epsilon_s=-2.692$, $\epsilon_d=-1.322$, and $\epsilon_p=-2.350$, in MeV.  It
also determines the structure coefficients $\alpha^\Gamma_{\gamma_1\gamma_2}$ of
the collective $S$, $D$, and $P$ pairs.  In principle these may vary with mass
number $A$ to reflect the changing structure of the collective pairs.  Here,
however, 
we derive (boson) operators completely from the two- and four-nucleon systems.
A strategy to account for the variation of the boson Hamiltonian with $A$ is
discussed below.  Actually, in the two-nucleon calculation there is no need to
introduce effective two-body operators since the eigenvalues and eigenvectors in
the restricted Hilbert space ${\mathbb H}_P$ do not differ from those in the
complete Hilbert space ${\mathbb H}$.

We thus derive the two-body interaction matrix elements between the bosons from
an analysis of the four-nucleon system.  The first step here is to diagonalize
the shell-model Hamiltonian in a complete two-pair basis ${\mathbb H}$,
following the procedure outlined in Sec.~\ref{s_npsm} to overcome the
non-orthogonality of this basis.  The resulting eigenspectrum should coincide
with the one obtained with any standard shell-model code, allowing a rigorous
check of the formalism and its implementation.  Next, we diagonalize the
Hamiltonian in the restricted Hilbert space ${\mathbb H}_P$, which is the
collective subspace defined in terms of the pairs derived from the two-nucleon
system.  We will report two different types of result:  (i) one with the
original ``bare'' shell-model Hamiltonian, and (ii) one with an effective
Hamiltonian, defined by the procedure outlined in Sec.~\ref{s_effop}.  The third
and final step is to use the mapping procedure of Sec.~\ref{s_map} to determine
the two-body part of the boson Hamiltonian. If $S$ and $D$ pairs are mapped onto
$s$ and $d$ bosons, without considering an isoscalar $P$ pair, then we call the
resulting boson model IBMb or IBMe, depending on whether the bare
or the effective shell-model Hamiltonian is used.  Likewise, if $S$, $D$, and
$P$ pairs are mapped onto $s$, $d$, and $p$ bosons, the resulting models are
referred to as \mbox{$p$-IBMb} or \mbox{$p$-IBMe}.  We emphasize that, unlike in
the two-nucleon system, for four nucleons a realistic shell-model Hamiltonian in
general couples the collective subspace to the rest of the space with
potentially important renormalization effects.  Therefore, the mapped two-body
matrix elements in IBMb and IBMe may differ significantly.

\begin{table*}[t]
\centering
\caption{Two-boson interaction matrix elements $\braket{ JT=0|\hat H^{\rm b}_2|JT=0}$
(in MeV), mapped from the KB3G interaction (\mbox{$p$-IBMb}, top)
and from the effective fermion interaction (\mbox{$p$-IBMe}, bottom).}
\label{t_v35-0}
\begin{ruledtabular}
\begin{tabular}{rrrrcrrrrcrr}
\multicolumn{4}{c}{$J=0$}&~~~~~~&\multicolumn{4}{c}{$J=2$}&~~~~~~&\multicolumn{2}{c}{$J=4$}\\
\cline{1-4}\cline{6-9}\cline{11-12} &$\ket{s^2}$&$\ket{d^2}$&$\ket{p^2}$&&
&$\ket{sd}$&$\ket{d^2}$&$\ket{p^2}$&& &$\ket{d^2}$\\
$\bra{ s^2}$&
$\begin{array}{r}-3.718\\[-3pt]-\mbox{\sl 3.909}\end{array}$&
$\begin{array}{r}-2.476\\[-3pt]-\mbox{\sl 2.093}\end{array}$&
$\begin{array}{r}3.792\\[-3pt]\mbox{\sl 4.418}\end{array}$&&
$\bra{ sd}$&
$\begin{array}{r}-6.078\\[-3pt]-\mbox{\sl 7.978}\end{array}$&
$\begin{array}{r}0.477\\[-3pt]-\mbox{\sl 0.601}\end{array}$&
$\begin{array}{r}-1.721\\[-3pt]-\mbox{\sl 1.195}\end{array}$&&
$\bra{ d^2}$&
$\begin{array}{r}-5.116\\[-3pt]-\mbox{\sl 8.642}\end{array}$\\
$\bra{ d^2}$&&
$\begin{array}{r}-3.436\\[-3pt]-\mbox{\sl 4.147}\end{array}$&
$\begin{array}{r}2.187\\[-3pt]\mbox{\sl 1.984}\end{array}$&&
$\bra{ d^2}$&&
$\begin{array}{r}-4.802\\[-3pt]-\mbox{\sl 7.898}\end{array}$&
$\begin{array}{r}0.826\\[-3pt]-\mbox{\sl 0.065}\end{array}$&&\\
$\bra{ p^2}$&&&
$\begin{array}{r}-0.620\\[-3pt]-\mbox{\sl 1.975}\end{array}$&&
$\bra{ p^2}$&&&
$\begin{array}{r}-0.327\\[-3pt]-\mbox{\sl 2.088}\end{array}$&&\\
\end{tabular}
\end{ruledtabular}
\end{table*}

\begin{table*}[t]
\centering
\caption{Two-boson interaction matrix elements $\braket{ JT=1|\hat H^{\rm
b}_2|JT=1}$ (in MeV), mapped from the KB3G interaction (\mbox{$p$-IBMb}, top)
and from the effective fermion interaction (\mbox{$p$-IBMe}, bottom).}
\label{t_v35-1}
\begin{ruledtabular}
\begin{tabular}{rrrrcrrrcrrr}
\multicolumn{4}{c}{$J=1$}&~~~~~~&\multicolumn{3}{c}{$J=2$}&~~~~~~&\multicolumn{3}{c}{$J=3$}\\
\cline{1-4}\cline{6-8}\cline{10-12} &$\ket{sp}$&$\ket{d^2}$&$\ket{dp}$&&
&$\ket{sd}$&$\ket{dp}$&& &$\ket{d^2}$&$\ket{dp}$\\
$\bra{ sp}$&
$\begin{array}{r}0.468\\[-3pt]\mbox{\sl 0.117}\end{array}$&
$\begin{array}{r}0.422\\[-3pt]\mbox{\sl 0.710}\end{array}$&
$\begin{array}{r}-1.032\\[-3pt]-\mbox{\sl 0.949}\end{array}$&&
$\bra{ sd}$&
$\begin{array}{r}-2.075\\[-3pt]-\mbox{\sl 2.518}\end{array}$&
$\begin{array}{r}-1.327\\[-3pt]-\mbox{\sl 1.990}\end{array}$&&
$\bra{ d^2}$&
$\begin{array}{r}-3.019\\[-3pt]-\mbox{\sl 4.649}\end{array}$&
$\begin{array}{r}0.861\\[-3pt]\mbox{\sl 0.559}\end{array}$\\
$\bra{ d^2}$&&
$\begin{array}{r}-2.328\\[-3pt]-\mbox{\sl 3.020}\end{array}$&
$\begin{array}{r}1.754\\[-3pt]\mbox{\sl 1.678}\end{array}$&&
$\bra{ dp}$&&
$\begin{array}{r}-1.441\\[-3pt]-\mbox{\sl 2.256}\end{array}$&&
$\bra{ dp}$&&
$\begin{array}{r}-0.435\\[-3pt]-\mbox{\sl 2.123}\end{array}$\\
$\bra{ dp}$&&&
$\begin{array}{r}-0.664\\[-3pt]-\mbox{\sl 1.267}\end{array}$&&&&&&&\\
\end{tabular}
\end{ruledtabular}
\end{table*}
The two-body matrix elements between the $s$ and $d$ bosons appear in
Table~\ref{t_v3}, both for the ``bare'' KB3G interaction (IBMb) and for its
effective version renormalized to the collective subspace (IBMe).  Because the
largest components of the $S$ and $D$ pairs are in the $1f_{7/2}$ shell, we can
compare the IBMe Hamiltonian with that obtained by Thompson {\it et
al.}~\cite{Thompson87} from a $1f_{7/2}$ shell-model interaction.  The bottom
row of Table \ref{t_v3} indeed shows that the IBMe Hamiltonian resembles the one
of Ref.~\cite{Thompson87}.  There are some differences, notably in the $J=0,T=2$
matrix elements, which suffer from a spurious $d^2$ state in the $1f_{7/2}$
mapping (a problem that is absent from the $pf$ mapping), and smaller
differences appear because the shell-model interactions and mapping procedures
(OAI {\it versus} democratic) are not exactly the same.  But overall, the
agreement is good.

The two-body matrix elements between the $s$, $d$, and $p$ bosons, both for the
\mbox{$p$-IBMb} and \mbox{$p$-IBMe}, are shown in Table~\ref{t_v35-0} for $T=0$
and in Table~\ref{t_v35-1} for $T=1$.  The $P$ pair does not influence the $T=2$
matrix elements, which therefore can be taken from Table~\ref{t_v3}.  A general
feature of the results, either with $s$ and $d$, or with $s$, $d$, and $p$
bosons, is that the diagonal matrix elements of the effective Hamiltonians are
more attractive (or less repulsive) than corresponding matrix elements of the
bare Hamiltonians.  This is to be expected because the renormalization takes
account of correlations from outside the collective subspace.

\begin{figure*}[t]
\begin{center}
\includegraphics[height=7cm]{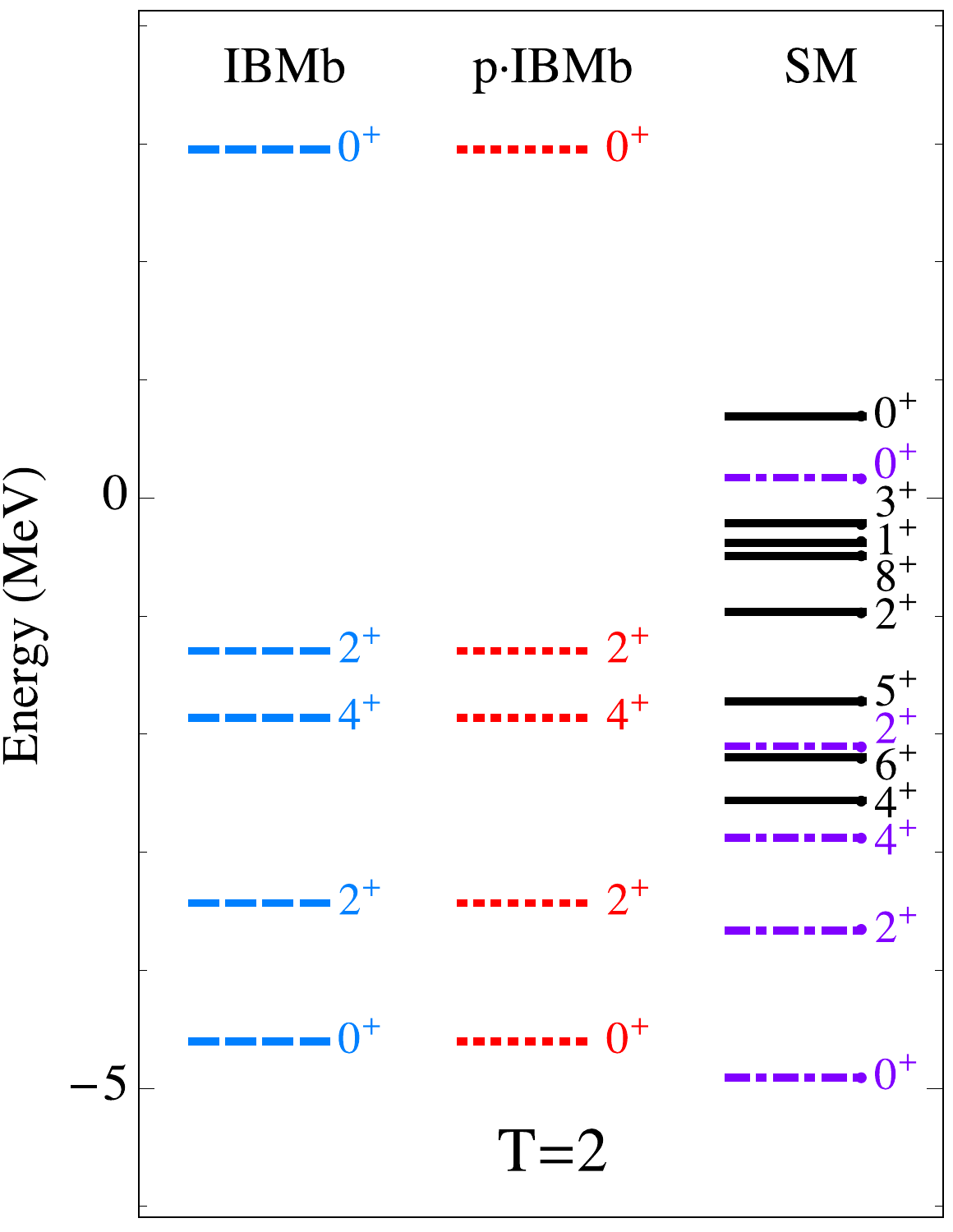}
\includegraphics[height=7cm]{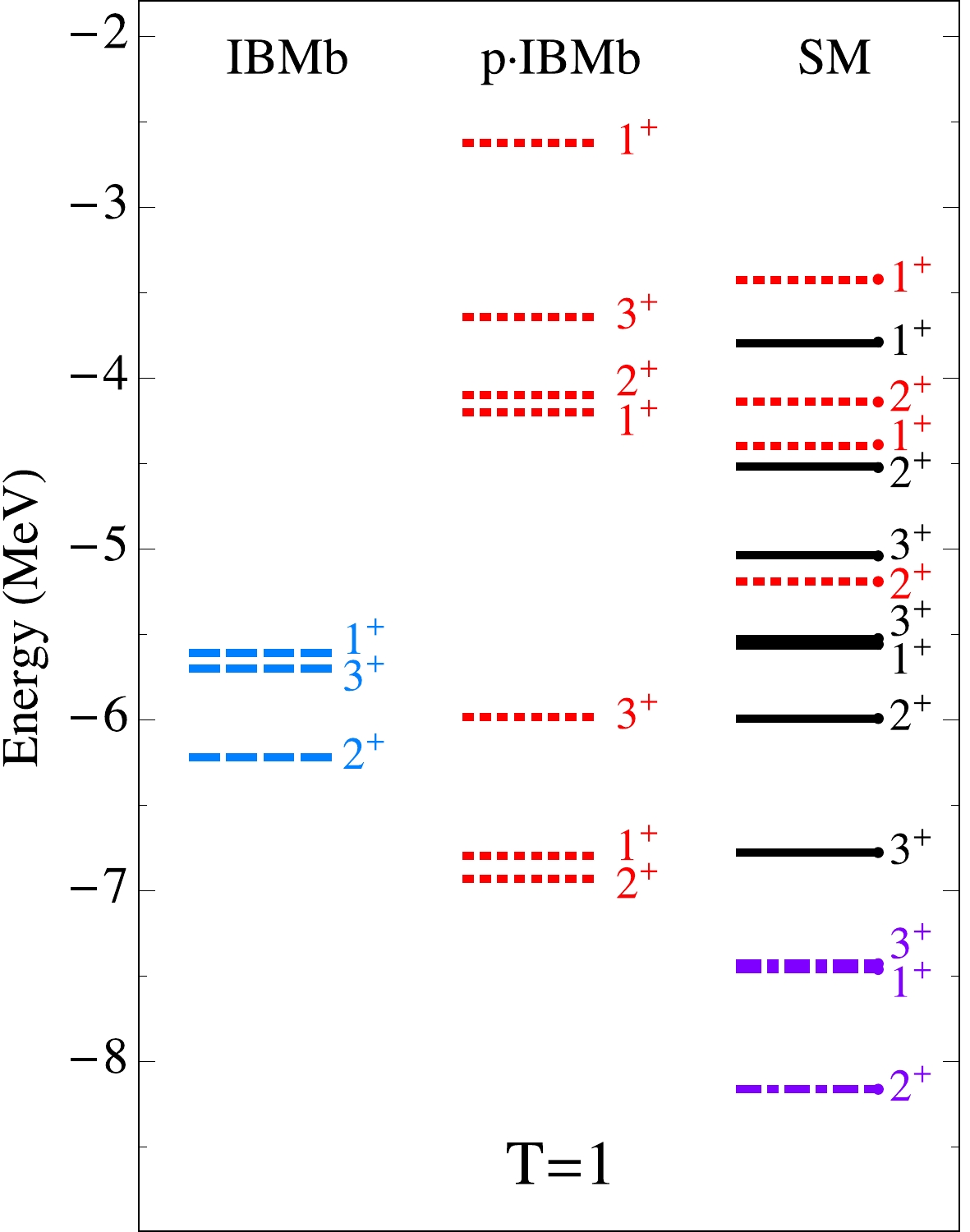}
\includegraphics[height=7cm]{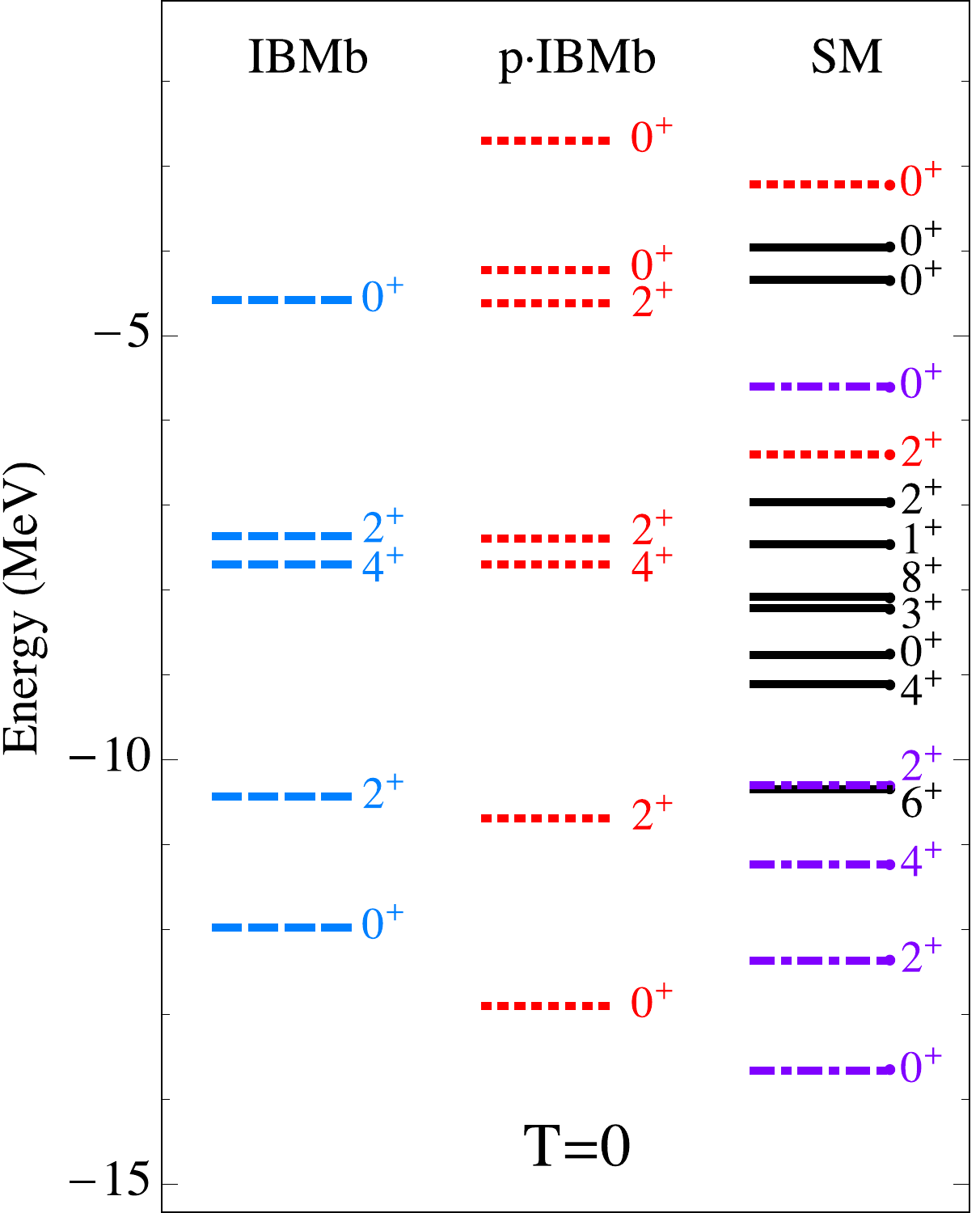}
\caption{ Four-nucleon spectra for $T=2$, $T=1$, and $T=0$, corresponding to
low-lying levels in the nuclei $^{44}$Ca, $^{44}$Sc, and $^{44}$Ti.  The left
column (dashed blue) shows all levels obtained with the bare Hamiltonian in the
collective subspace constructed from $S$ and $D$ pairs.  The middle column
(dotted red) shows the same for $S$, $D$, and $P$ pairs.  The right column shows
the low-energy levels obtained in the shell model with the KB3G interaction; the
ones that are exactly reproduced with an effective Hamiltonian in the $SD$ and
$SDP$ subspaces are drawn in dash-dotted purple and those that are reproduced
only in the $SDP$ subspace in dotted red.}
\label{f_a44}
\end{center}
\end{figure*}

Figure~\ref{f_a44} shows four-nucleon spectra for $T=2$, $T=1$, and $T=0$,
corresponding to low-lying levels in the nuclei $^{44}$Ca, $^{44}$Sc, and
$^{44}$Ti.  In each case the figure shows the levels calculated in the shell
model (SM) with the KB3G interaction.  Some of these levels are exactly
reproduced with an effective Hamiltonian constructed for a particular subspace:
SM levels in dashed blue pertain to the $SD$ subspace and those in dotted red to the $SDP$
subspace, while SM levels in dash-dotted purple are calculated in both
subspaces.  For comparison, the figure also shows the results produced by the
bare KB3G Hamiltonian in the two collective subspaces.

All levels in Fig.~\ref{f_a44} result from {\em fermionic} calculations, in
which the Pauli principle is fully taken into account, though possibly in a
truncated Hilbert space.  The various fermionic systems are mapped onto
corresponding bosonic systems, consisting of either $s$ and $d$ bosons, or $s$,
$d$, and $p$ bosons.  For a four-nucleon system and a boson Hamiltonian
containing up to two-body interactions between the bosons, the mapping is exact.
Therefore, the levels in the left (middle) columns of Fig.~\ref{f_a44} are also
obtained in the {\em boson} calculation with the bare Hamiltonian of the
IBMb (\mbox{$p$-IBMb}) while the colored levels in the right column are
obtained in IBMe (\mbox{$p$-IBMe}).

In summary, the two-boson calculations (with the IBMe or \mbox{p-IBMe}
Hamiltonians) {\em exactly} reproduce the energy of some
eigenstates of the four-nucleon shell-model calculation.  As explained in
Sec.~\ref{s_effop}, the normal procedure is to select those that have maximal
overlap with the eigenstates in the two-pair basis.  This set usually includes
the yrast state but not necessarily the yrare state.  For example, the shell
model gives a $J=0,T=0$ ground state at $-13.668$~MeV, which we also include in
the \mbox{$(p)$-IBMe}; the next $J=0,T=0$ state in the two-boson calculation is
at $-5.601$~MeV and corresponds to the {\em third} shell-model state with those
quantum numbers.  For the $J=4$ states, both with $T=0$ and $T=2$, we find that
the $D^2$ pair state is fragmented over the yrast and yrare shell-model states.
We choose to assign the boson state to the lowest one, irrespective of the
overlap; this choice conforms to the one of Thompson {\it et al.}~\cite{Thompson87}.

From Fig.~\ref{f_a44} it is apparent that the dash-dotted purple SM levels are concentrated in
the low-energy region, indicating that the $s$ and $d$ bosons capture the
essential collective degrees of freedom.  The same statement cannot be made
about the dotted red levels, which also occur at higher energies.  This is a first
indication that the isoscalar $p$ boson is not crucial for describing low-lying
spectra, a fact that is not surprising in light of past work~\cite{Afanasjev12}.

\subsection{Mapping of the $0\nu\beta\beta$-decay operator}
\label{ss_mapbb}
The boson $0\nu\beta\beta$-decay operator ($0\nu\beta\beta$ operator for short)
is also determined by the $A=42$ and $A=44$ nuclei.  Although the
$0\nu\beta\beta$ operator has no fermion one-body term, its boson one-body terms
are non-zero and determined by the analog of Eq.~\eqref{e_hme1}.  The
shell-model $0\nu\beta\beta$ operator, together with the KB3G interaction
applied to $A=42$, leads to $\braket{ s\|\hat T^{\rm
b}_{1,\beta\beta}\|s}=-11.395$ and $\braket{ d\|\hat T^{\rm
b}_{1,\beta\beta}\|d}=-15.179$.  No $0\nu\beta\beta$ transition occurs between
$P$ pairs with $T=0$, and hence $\braket{ p\|\hat T^{\rm
b}_{1,\beta\beta}\|p}=0$.  The two-body part of the $0\nu\beta\beta$
operator is specified by the reduced matrix elements $\braket{b_1b_2;JT_{\rm
f}\|\hat T^{\rm b}_{2,\beta\beta}\|b'_1b'_2;JT_{\rm i}}$.  For the $T_{\rm
i}=2\rightarrow T_{\rm f}=0$, $T_{\rm i}=1\rightarrow T_{\rm f}=1$, and $T_{\rm
i}=2\rightarrow T_{\rm f}=0$ transitions the $p$ boson can contribute while the
$T_{\rm i}=2\rightarrow T_{\rm f}=2$ transitions are independent of the $p$
boson.

The total $0\nu\beta\beta$ operator, both in the IBMb and \mbox{$p$-IBMb}, and
their effective versions, the IBMe and \mbox{$p$-IBMe}, is completely specified
by the reduced matrix elements $\braket{b_1b_2;JT_{\rm f}\|\hat T^{\rm
b}_{2,\beta\beta}\|b'_1b'_2;JT_{\rm i}}$.  Of course, similar mappings can be
executed for separate pieces of the $0\nu\beta\beta$ operator, such as its
Gamow-Teller part.  The effective-operator theory of Suzuki and
Lee~\cite{suzuki80} ensures that the transition matrix elements between
eigenstates in the restricted Hilbert space ${\mathbb H}_P$ coincide {\em
exactly} with those between some of the eigenstates in the complete Hilbert
space ${\mathbb H}$.

We conclude this and the previous subsection by re-emphasizing that the
formalism developed in this paper allows us to derive a boson Hamiltonian and,
in general, boson operators that exactly reproduce the properties of a subset of
the shell-model eigenstates of all two- and four-nucleon systems.

\subsection{Results for the energies}
\label{ss_resen}
We now turn to systems with more nucleons and consider nuclei for which a
shell-model calculation is feasible in the complete Hilbert space, in order to
compare its results with those of the IBM.  Our procedure incorporates no
$A$ dependence into the IBM operators, so that we have neither the
mass-dependent structure coefficients $\alpha^\Gamma_{\gamma_1\gamma_2}$
mentioned earlier nor a dependence of the IBM Hamiltonian on the
boson number $n$ and isospin $T$, which is discussed in
Refs.~\cite{Evans93,Evans95}.  Not only is it difficult to combine the two
effects but in addition the ($n,T$)-dependence as derived in
Refs.~\cite{Evans93,Evans95} applies only to a seniority-based mapping, the
generalization of which to an arbitrary system of bosons is not obvious.  For
the purpose of this paper, therefore, we propose the following heuristic method
to obtain an $A$-dependent IBM Hamiltonian and, in general, $A$-dependent
IBM operators.

\begin{figure*}[t]
\begin{center}
\includegraphics[height=7cm]{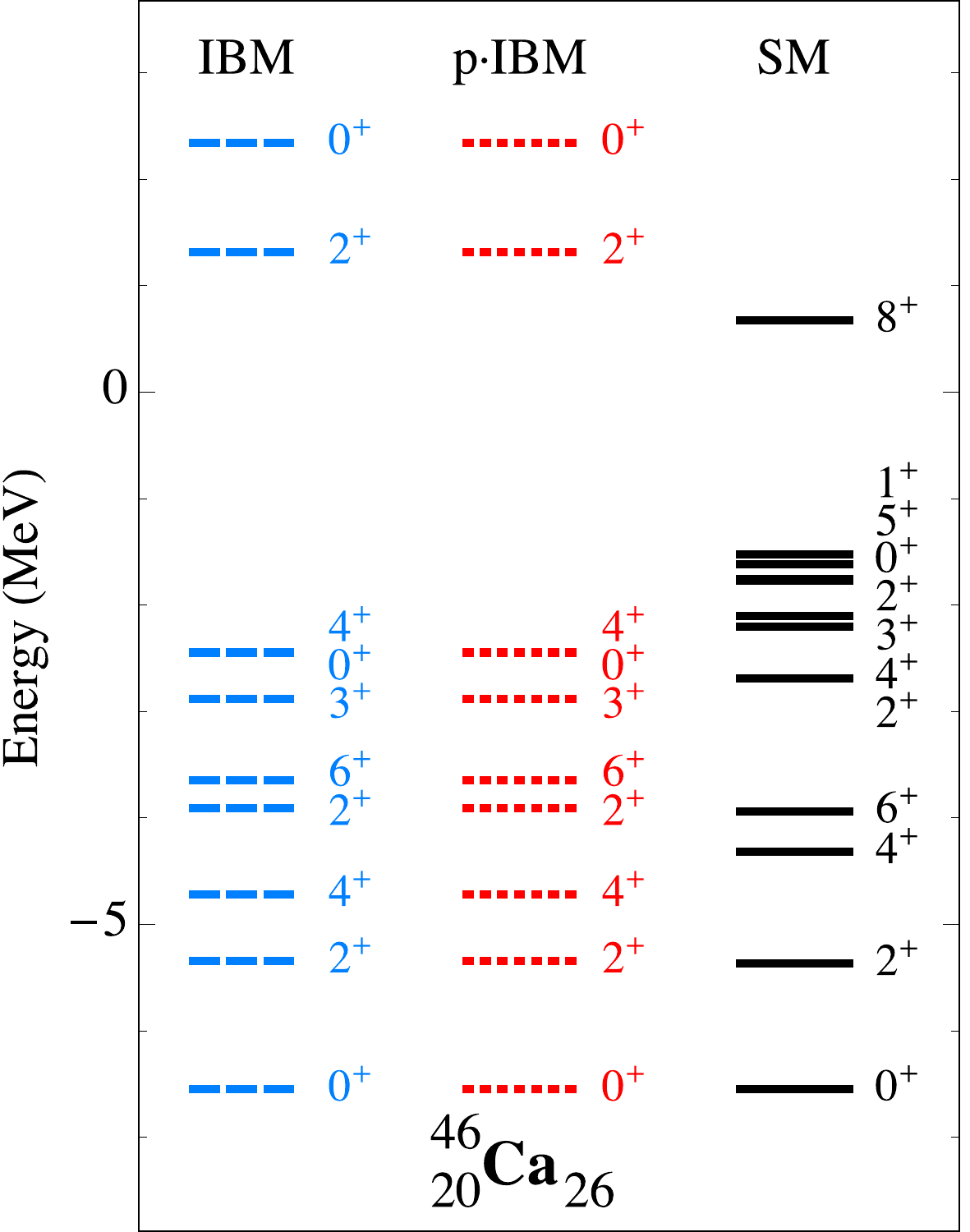}
\includegraphics[height=7cm]{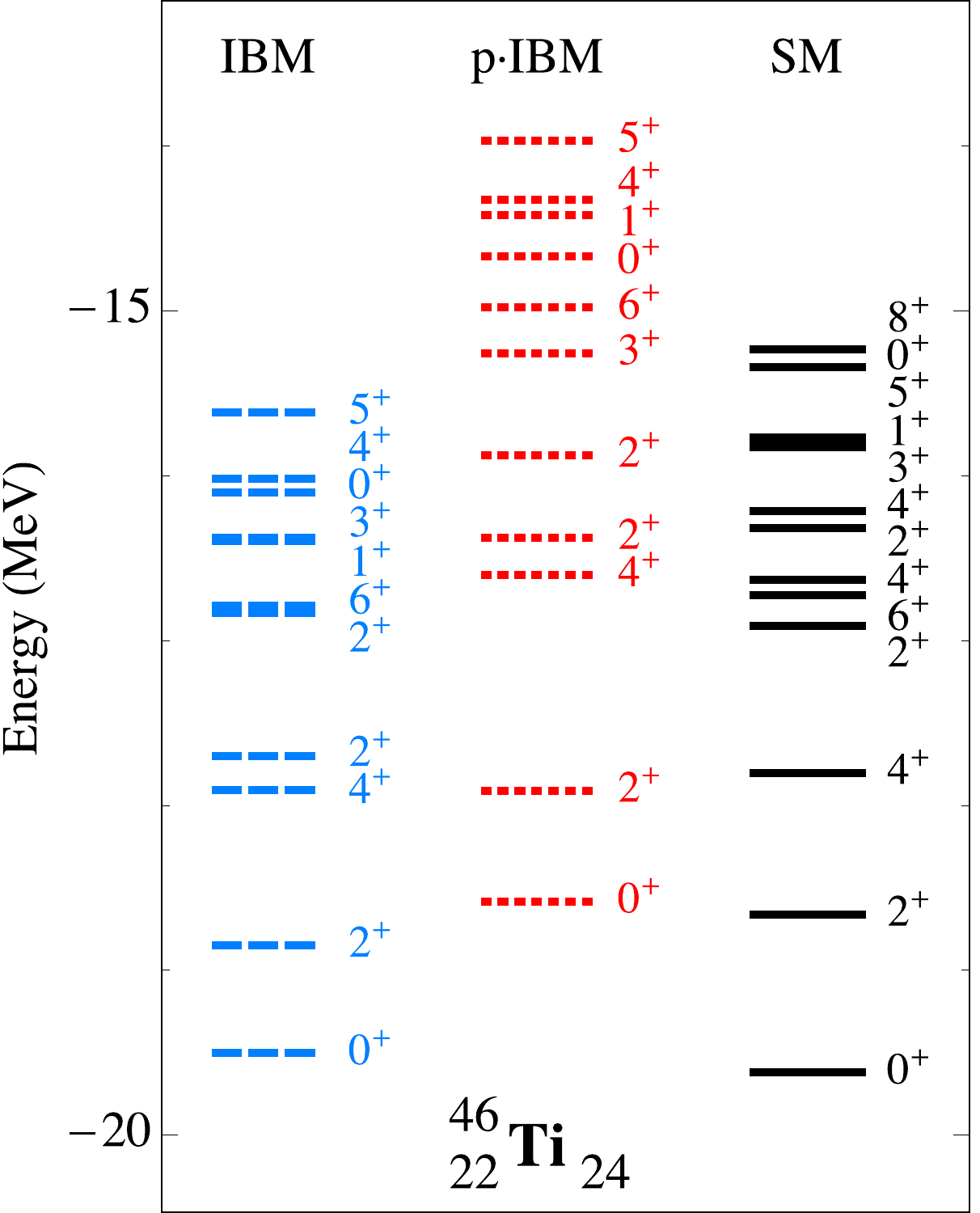}
\caption{ Spectra of the $A=46$ nuclei $^{46}$Ca ($T=3$) and $^{46}$Ti ($T=1$).
The shell-model spectra (SM) are produced by the KB3G interaction with six
nucleons in the $pf$ shell.  The IBM spectra, with three bosons ($sd$ or $sdp$),
are produced by a Hamiltonian interpolated between the bare $\hat H^{\rm b}_{\rm
b}$ and the effective $\hat H^{\rm b}_{\rm e}$ (see text).}
\label{f_a46}
\end{center}
\end{figure*}

As explained in Subsec.~\ref{ss_maph}, for a given bosonic system (e.g.,
$sd$ or $sdp$) the mapping defines a bare boson Hamiltonian
$\hat H^{\rm b}_{\rm b}$---obtained from the bare fermion Hamiltonian---as
well as an effective one $\hat H^{\rm b}_{\rm e}$.  
The former underbinds the two-boson system, when compared with the shell-model
result for four nucleons, while the latter exactly reproduces the shell-model
binding energy.  In the use of these Hamiltonians for systems with more bosons,
we have consistently found that, for a given boson number $n$, angular momentum
$J$, and isospin $T$, $\hat H^{\rm b}_{\rm b}$ underestimates the corresponding
(positive) shell-model binding energy and $\hat H^{\rm b}_{\rm e}$ overestimates
it.  Therefore, although we have no formal proof of it, we conjecture the
following inequalities:
\begin{equation}
\braket{\hat H^{\rm b}_{\rm e}}_{n,J,T}\leq
\braket{\hat H^{\rm f}}_{2n,J,T}\leq
\braket{\hat H^{\rm b}_{\rm b}}_{n,J,T}\,,
\label{e_ineq}
\end{equation}
where $\braket{\hat H^{\rm f}}_{2n,J,T}$ is the lowest eigenvalue, for a given
nucleon number $2n$, angular momentum $J$, and isospin $T$, of the shell-model
Hamiltonian in the complete Hilbert space ${\mathbb H}$.  These inequalities
suggest the use of an ($n,T$)-dependent boson Hamiltonian of the form
\begin{equation}
\hat H^{\rm b}=x\hat H^{\rm b}_{\rm b}+(1-x)\hat H^{\rm b}_{\rm e}\,,
\label{e_interpol}
\end{equation}
with $x$ an ($n,T$)-dependent parameter between 0 and 1 that we consider
adjustable, to be determined by a comparison with the spectrum of the
shell-model Hamiltonian in the complete Hilbert space ${\mathbb H}$.  By
construction $x=0$ for $n=2$ bosons and we expect $x$ to increase with
increasing $n$ and $T$.

Figure~\ref{f_a46} shows spectra of nuclei with mass number $A=46$; the results
of the interpolation procedure can be called satisfactory.
The panels in the figure are labeled with the nuclei and spectra refer to their
low-energy levels with isospin $T=|T_z|$.  Since isospin symmetry is conserved
in both the shell model and the IBM, the calculated spectra in $^{46}$Ca and
$^{46}$Ti are identical to those of the mirror nuclei $^{46}$Fe and $^{46}$Cr.
(Because both the shell model and the IBM produce absolute energies, one would
need different Coulomb corrections in the mirror nuclei.) The levels in
$^{46}$Ca have isospin $T=3$ and are not affected by the $p$ bosons; the
$^{46}$Ca spectra in IBM and \mbox{$p$-IBM} are consequently identical.  For the
$T=1$ levels of $^{46}$Ti, on the other hand, the IBM and \mbox{$p$-IBM} yield
different results.  In the IBM a value of $x$ can be chosen such that the
binding energies {\em and} the excitation spectra are reasonably well
reproduced.  That is not the case in the \mbox{$p$-IBM}: If $x$ is adjusted to
reproduce the shell-model binding energy of $^{46}$Ti, then an unrealistic
excitation spectrum results, with a $0^+$-$2^+$ energy splitting that is far too
low.  This difficulty confirms our suspicion that the isoscalar $p$ boson does
not play a vital role in the spectroscopy of light $pf$-shell nuclei.  Not only
does it render the mapping to the bosonic system more complex but it also
worsens the results of the simpler IBM.  We have, however, yet to examine its
role in $0\nu\beta\beta$ decay.  

\begin{figure*}[t]
\begin{center}
\includegraphics[height=7cm]{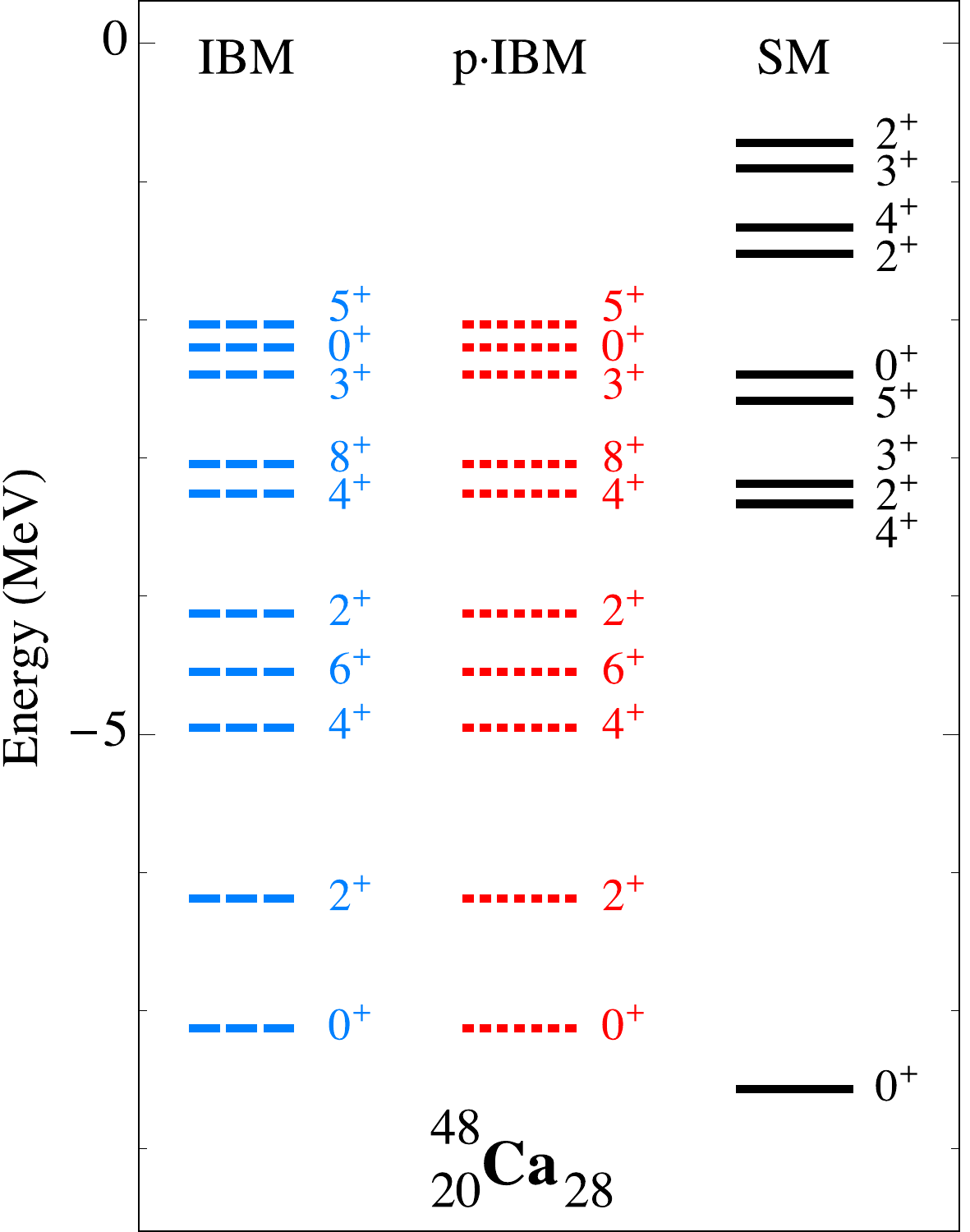}
\includegraphics[height=7cm]{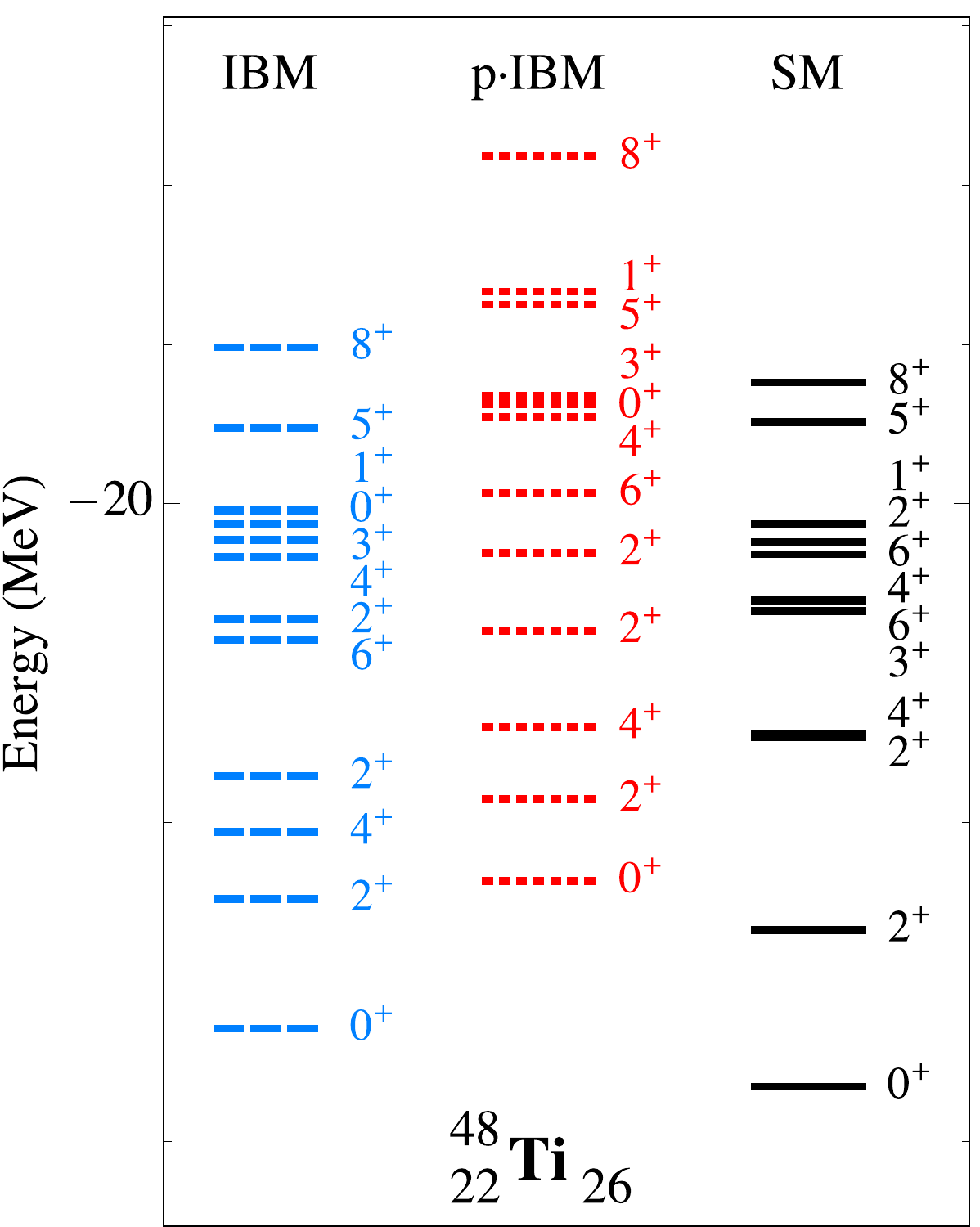}
\includegraphics[height=7cm]{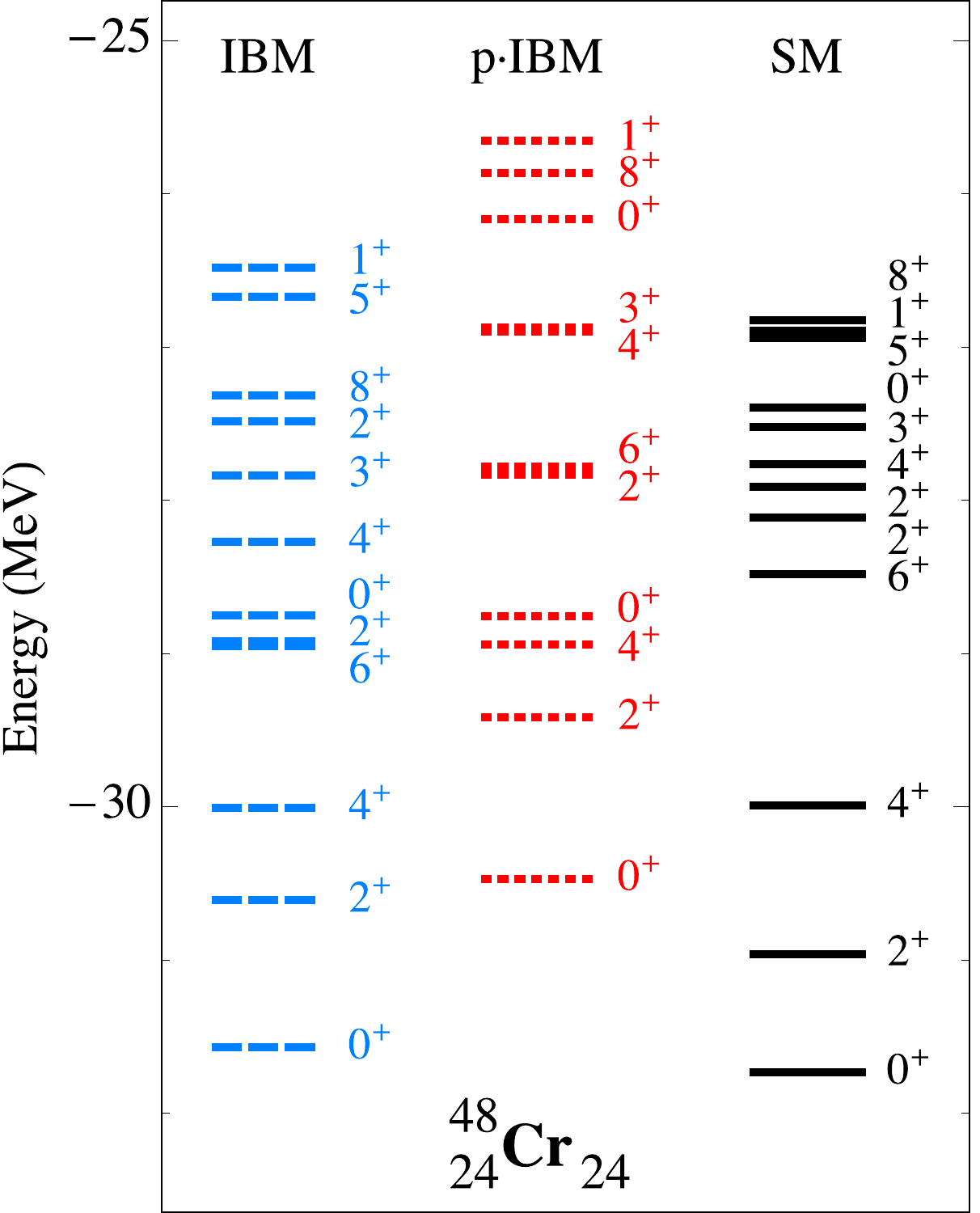}
\caption{ Spectra of the $A=48$ nuclei $^{48}$Ca ($T=4$), $^{48}$Ti ($T=2$), and
$^{48}$Cr ($T=0$).  The shell-model spectra (SM) are produced by the KB3G
interaction with eight nucleons in the $pf$ shell.  The IBM spectra, with four
bosons ($sd$ or $sdp$), are produced by a Hamiltonian interpolated between the
bare $\hat H^{\rm b}_{\rm b}$ and the effective $\hat H^{\rm b}_{\rm
e}$ (see text).}
\label{f_a48}
\end{center}
\end{figure*}
Figure~\ref{f_a48} shows spectra for nuclei with mass number $A=48$.  The boson
approximation clearly breaks down in $^{48}$Ca, a fact that is unsurprising
because the sub-shell closure at neutron number $N-28$ should cause the
structure of the collective pairs to change dramatically from we constructed in
$A=42$ nuclei.  The binding energy of the $^{48}$Ca ground state is not badly
wrong, however, and that particular state, which consists mostly of eight
neutrons in the $f_{7/2}$ shell but also includes correlations from the
$p_{3/2}$, $f_{5/2}$, and $p_{1/2}$, may still be described well enough to use
in calculating the $0\nu\beta\beta$ matrix element.  The spectra for $A=48$, however, again
confirm our statement $p$ bosons do not improve spectra.

\begin{figure*}[t]
\begin{center}
\includegraphics[height=7cm]{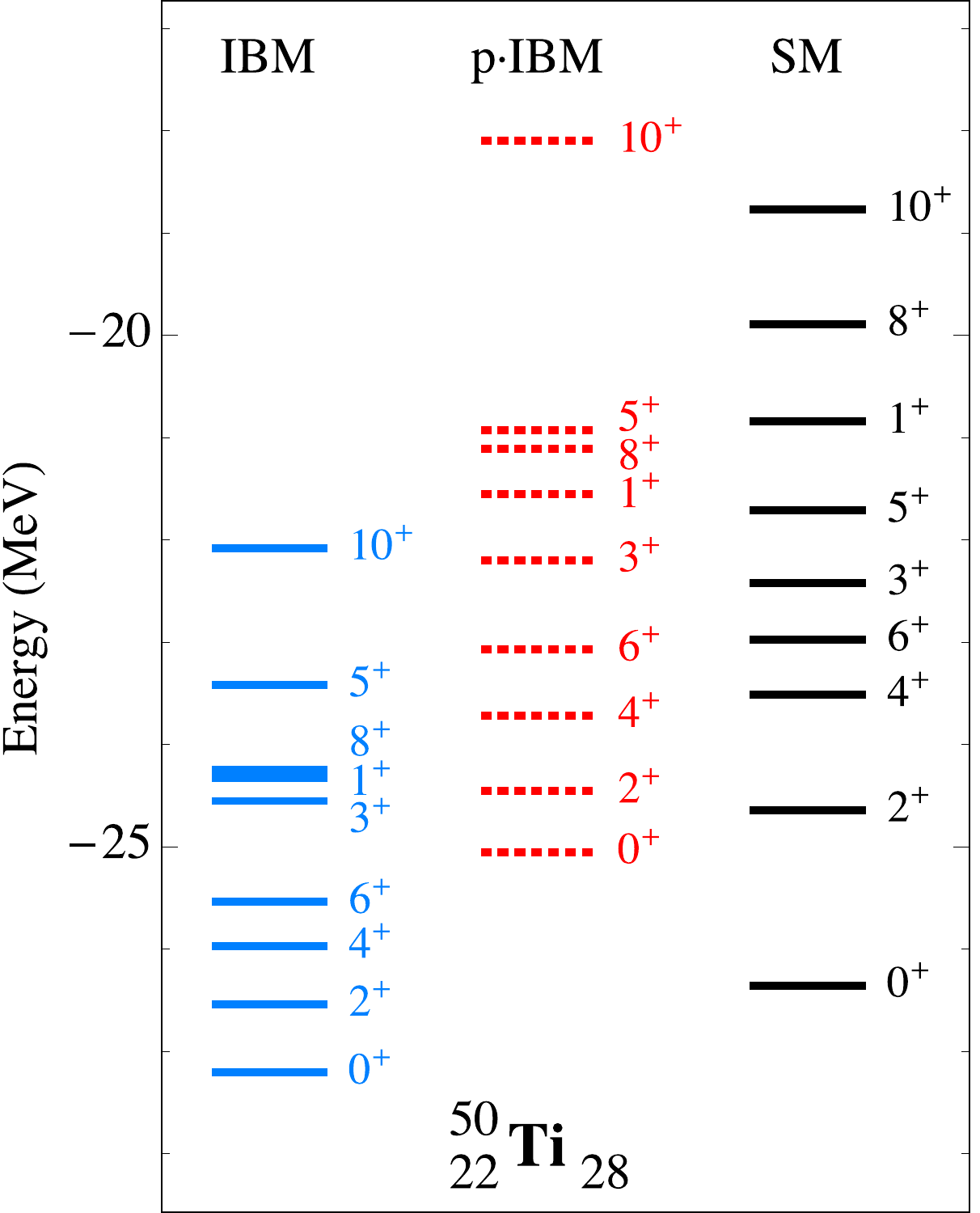}
\includegraphics[height=7cm]{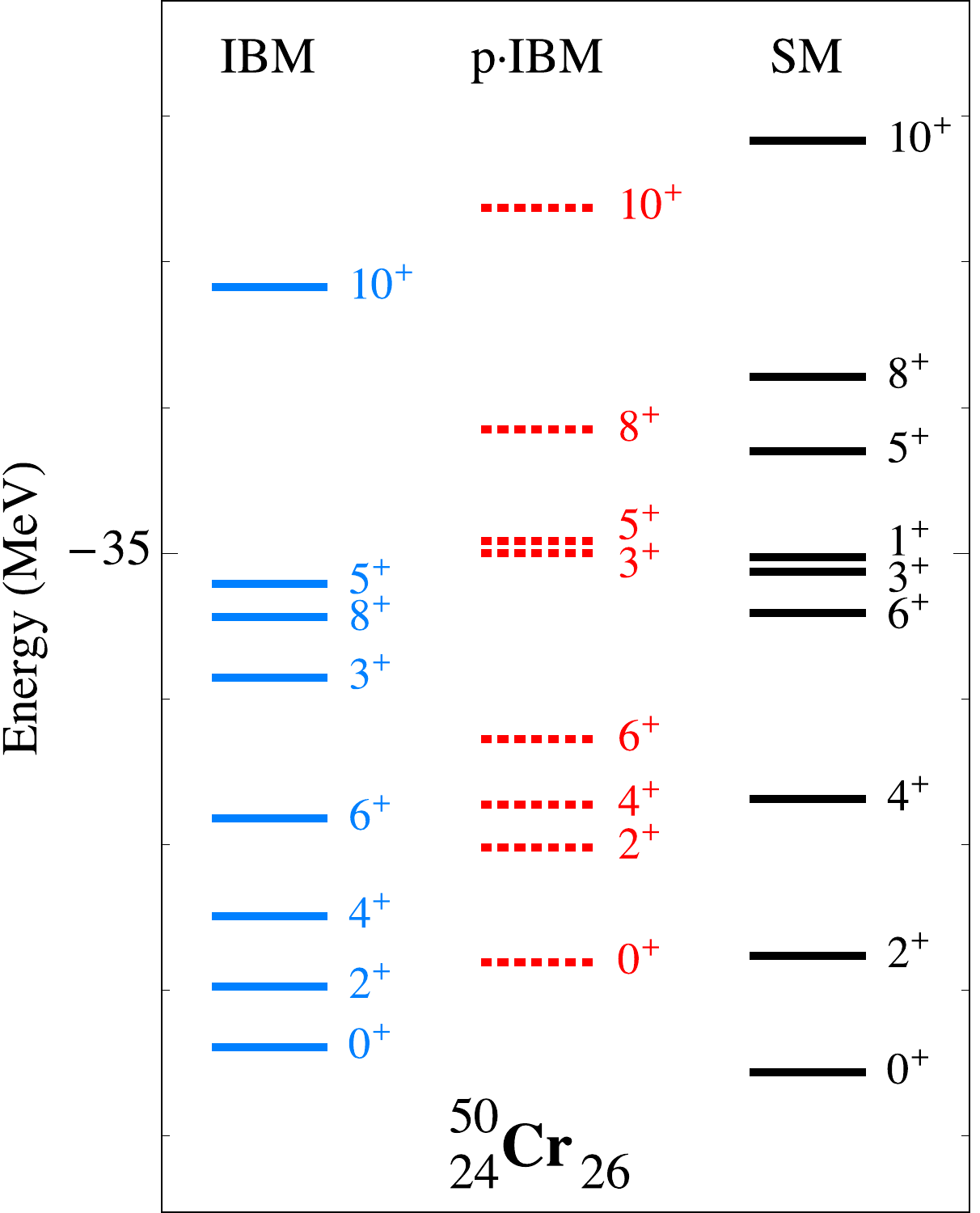}
\caption{ Spectra of the $A=50$ nuclei $^{50}$Ti ($T=3$) and $^{50}$Cr ($T=1$).
The shell-model spectra (SM) are produced by the KB3G interaction with ten
nucleons in the $pf$ shell.  The IBM spectra, with five bosons ($sd$ or $sdp$),
are produced by a Hamiltonian interpolated between the bare $\hat H^{\rm b}_{\rm
b}$ and the effective $\hat H^{\rm b}_{\rm e}$ (see text).}
\label{f_a50}
\end{center}
\end{figure*}
Finally, Fig.\ \ref{f_a50} shows spectra for nuclei with mass number $A=50$.
The neutron sub-shell closure at $N=28$ causes problems again in $^{50}$Ti.
Unlike $^{48}$Ca, for which only $T=2$ matrix elements enter the boson
calculation, $^{50}$Ti has ground-state structure that depends on {\em all}
boson matrix elements, including those with $T=1$ and $T=0$.  The many
significant matrix elements might lead the IBM and \mbox{$p$-IBM} to
overestimate the degree of correlation in the $^{50}$Ti ground state.  The same
seems to be true in the IBM without $p$ bosons for $^{50}$Cr.  We could improve
the excitation energies in these nuclei, if we wanted, by relaxing the
requirement of matching the binding energies.

\subsection{Results for $0\nu\beta\beta$-decay transitions}
\label{ss_resbb}
We turn finally to $0\nu\beta\beta$ matrix elements.  Our main interest at this
point is a comparison of the results of the shell model, as reported by
Men\'endez {\it et al.}~\cite{men16}, with those of the IBM and \mbox{$p$-IBM}.
The matrix elements depend on the values of the Hamiltonian-interpolation
parameter $x$ in the initial and final nuclei.  We can also assign a similar
parameter $x_{\beta\beta}$ to the $0\nu\beta\beta$ operator, that is we can use
a linear combination of the bare and effective $0\nu\beta\beta$ operators the
same way we do for the Hamiltonian in Eq.\ \eqref{e_interpol}.  Here we make the
simplest choice for $x_{\beta\beta}$, setting it equal to the average of the
Hamiltonian $x$ parameters in the initial and final nuclei.

\begin{figure}
\begin{center}
\includegraphics[width=\columnwidth]{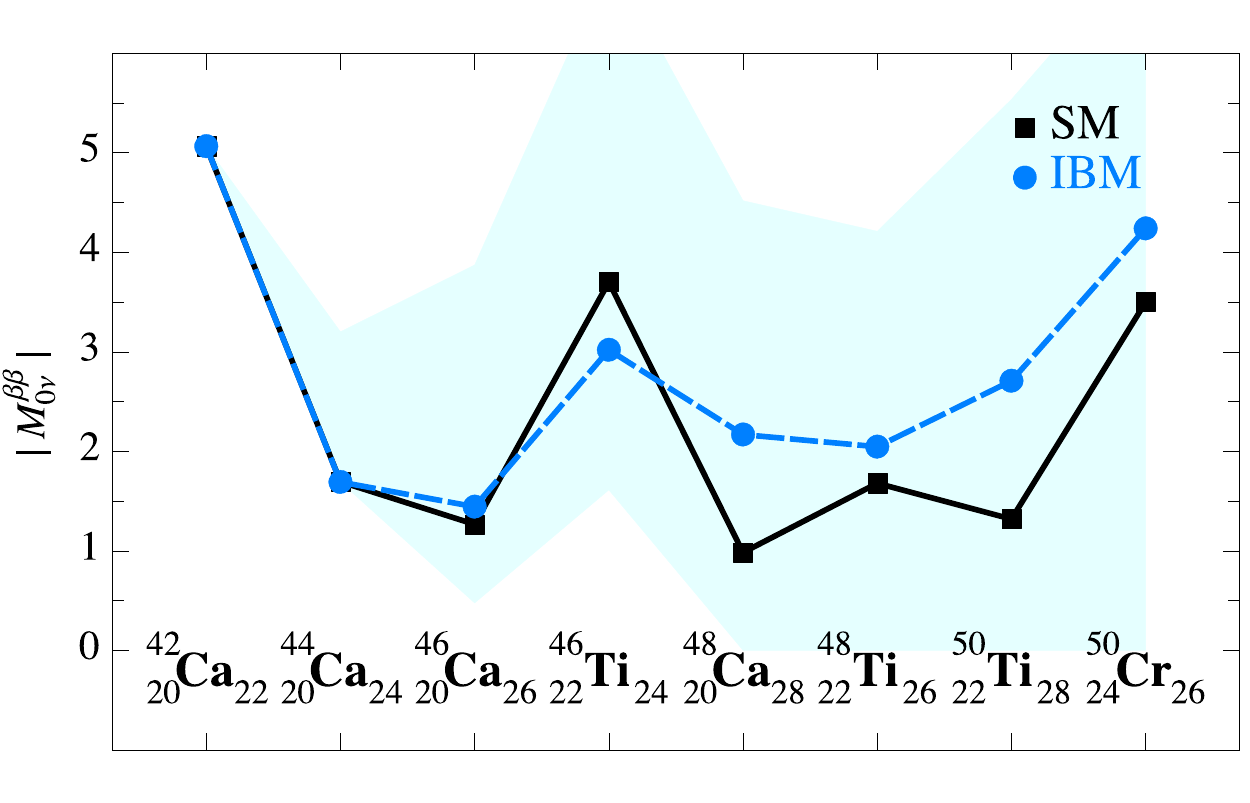}
\includegraphics[width=\columnwidth]{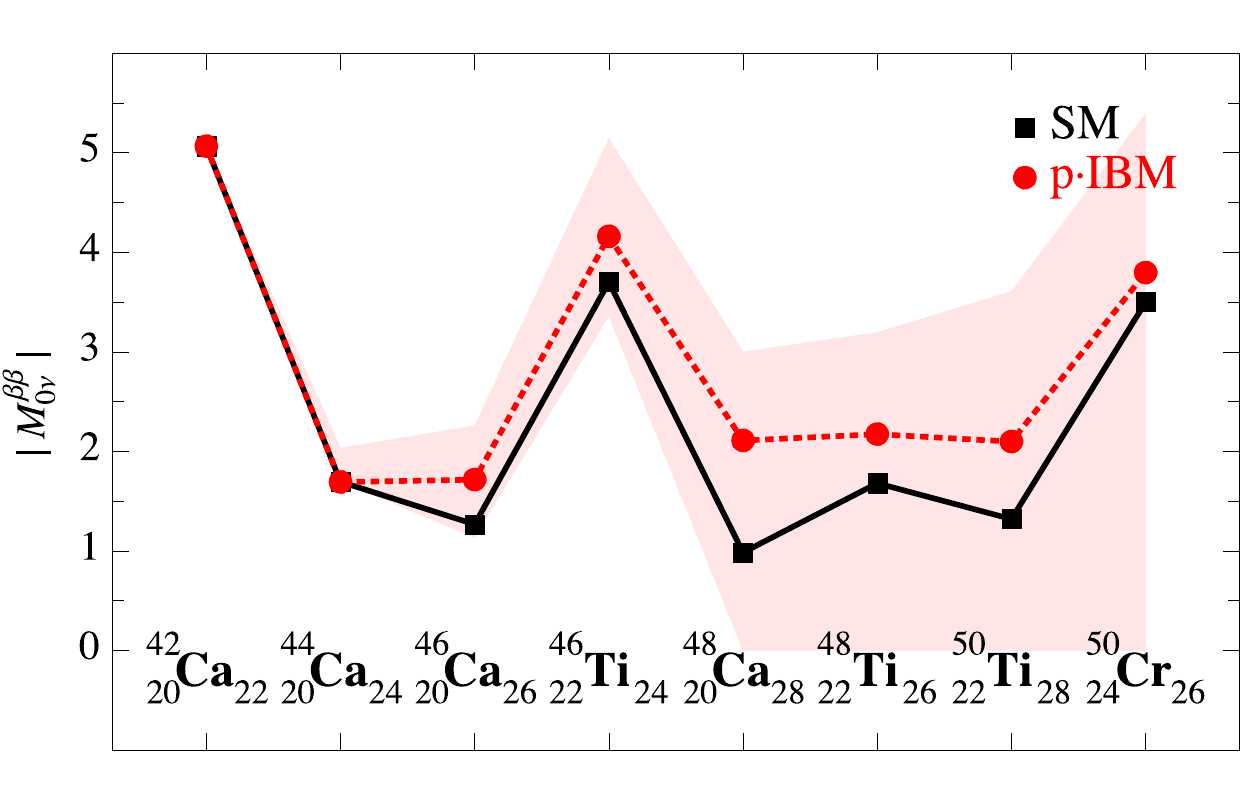}
\caption{ Eight $0\nu\beta\beta$ $0^+_1\rightarrow0^+_1$ matrix elements between
$f_{7/2}$-shell nuclei, calculated in the shell model~\cite{men16} and in IBM
(top) and \mbox{$p$-IBM} (bottom).  The full black lines connect the shell-model
results and the dashed blue (dotted red) lines the IBM (\mbox{$p$-IBM}) results
with an $x$ parameter fit to energy spectra (see text).  The shaded areas
indicate upper and lower limits defined by $x=1$ (bare operators) and $x=0$
(effective operators).}
\label{f_b2}
\end{center}
\end{figure}
\begin{figure}
\begin{center}
\includegraphics[width=\columnwidth]{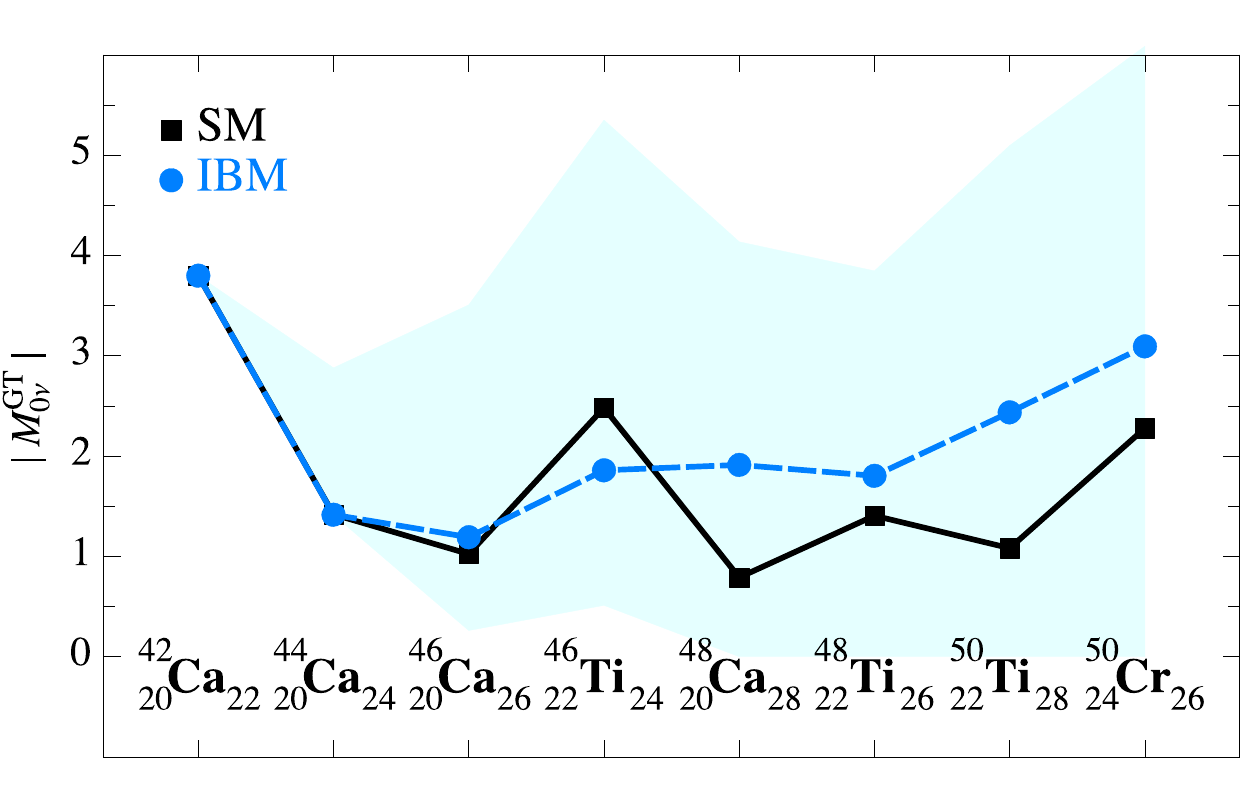}
\includegraphics[width=\columnwidth]{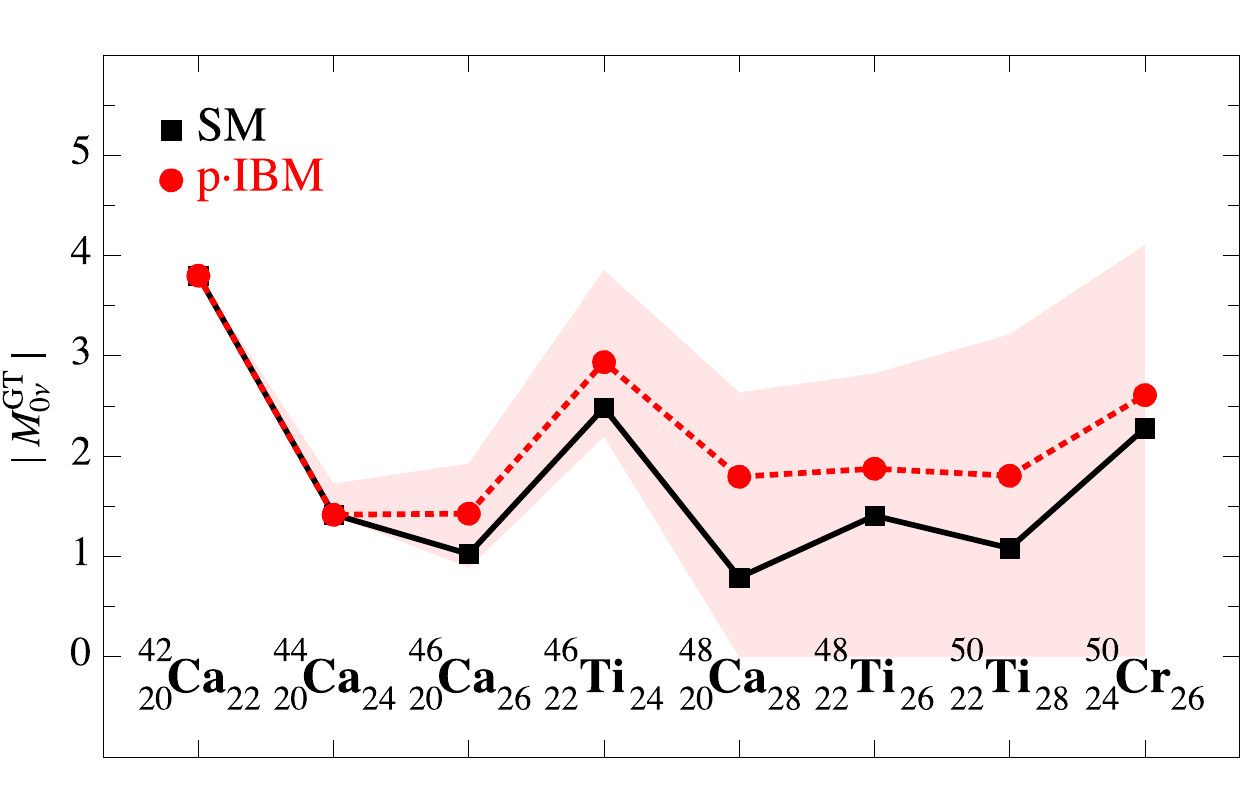}
\caption{ Same as Fig.~\ref{f_b2} for the Gamow-Teller part of eight
$0\nu\beta\beta$ $0^+_1\rightarrow0^+_1$ matrix elements between $f_{7/2}$-shell
nuclei.}
\label{f_b2gt}
\end{center}
\end{figure}

\begin{figure*}
\includegraphics[width=\columnwidth]{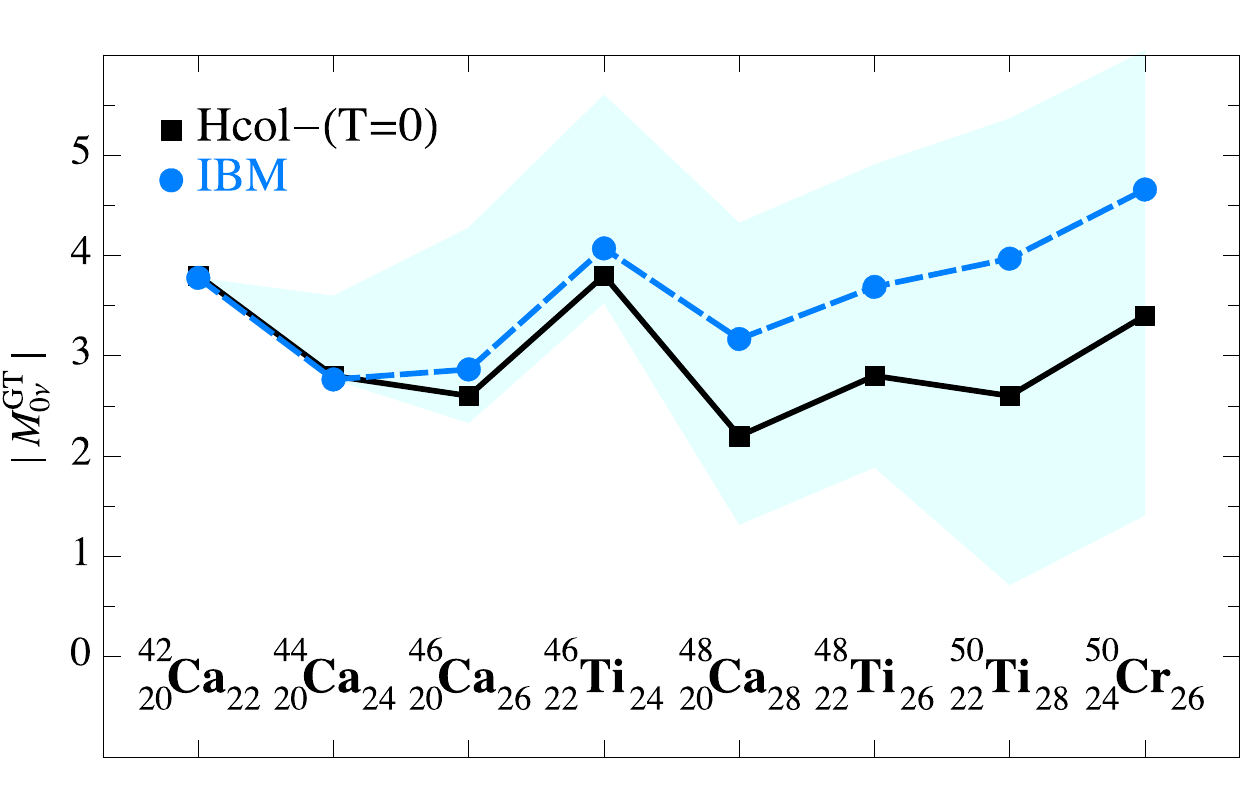}
\includegraphics[width=\columnwidth]{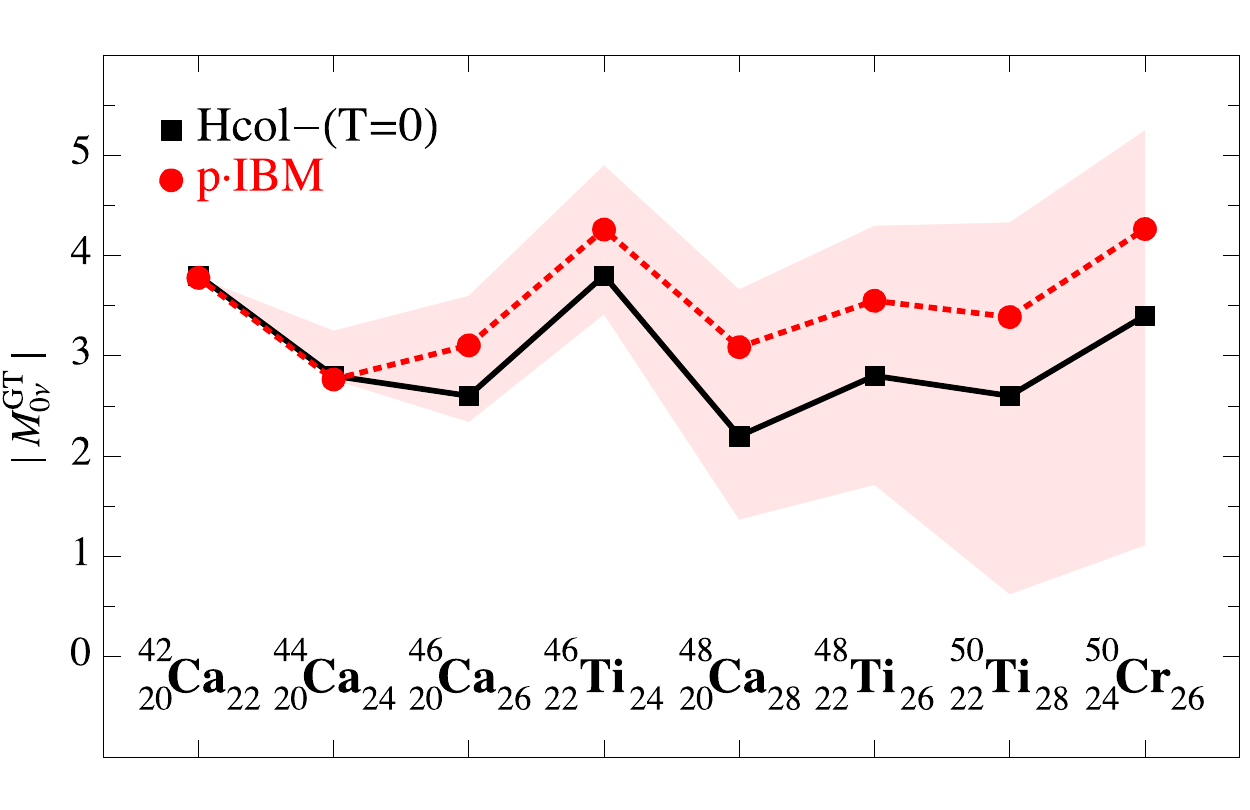}
\includegraphics[width=\columnwidth]{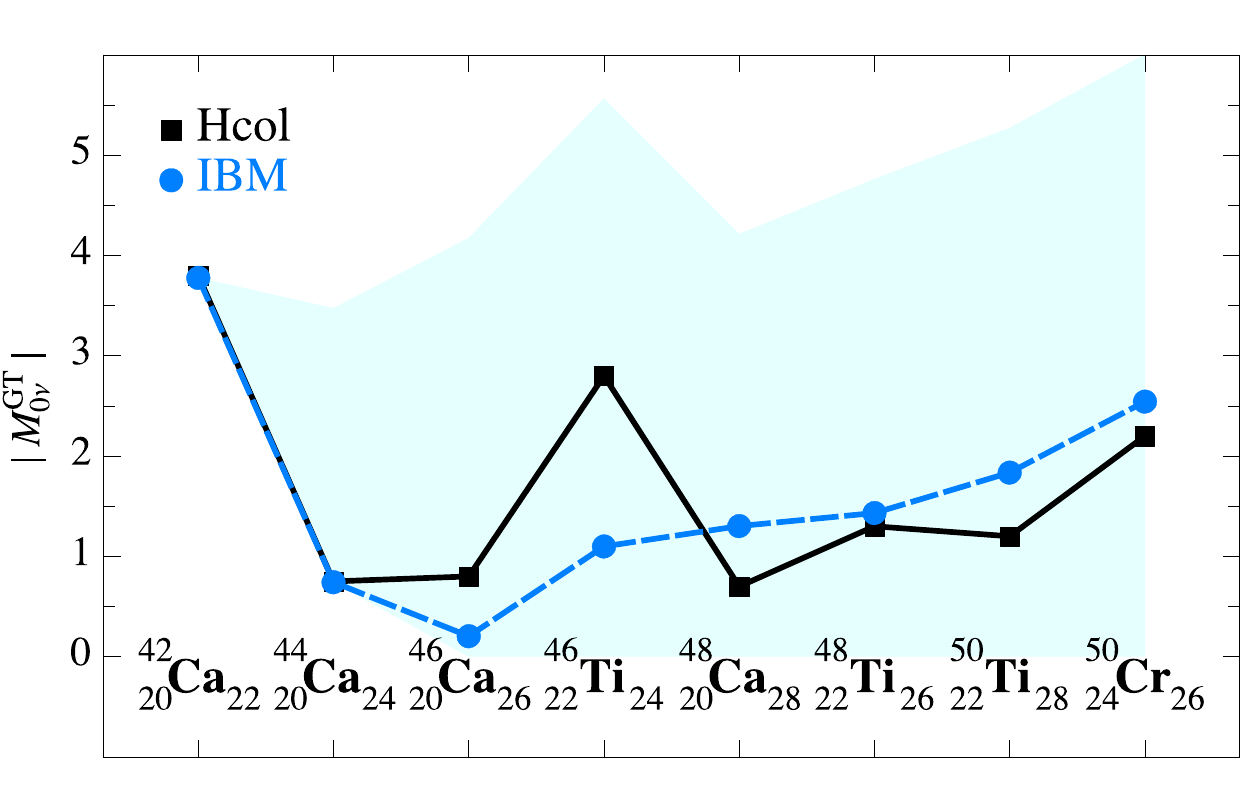}
\includegraphics[width=\columnwidth]{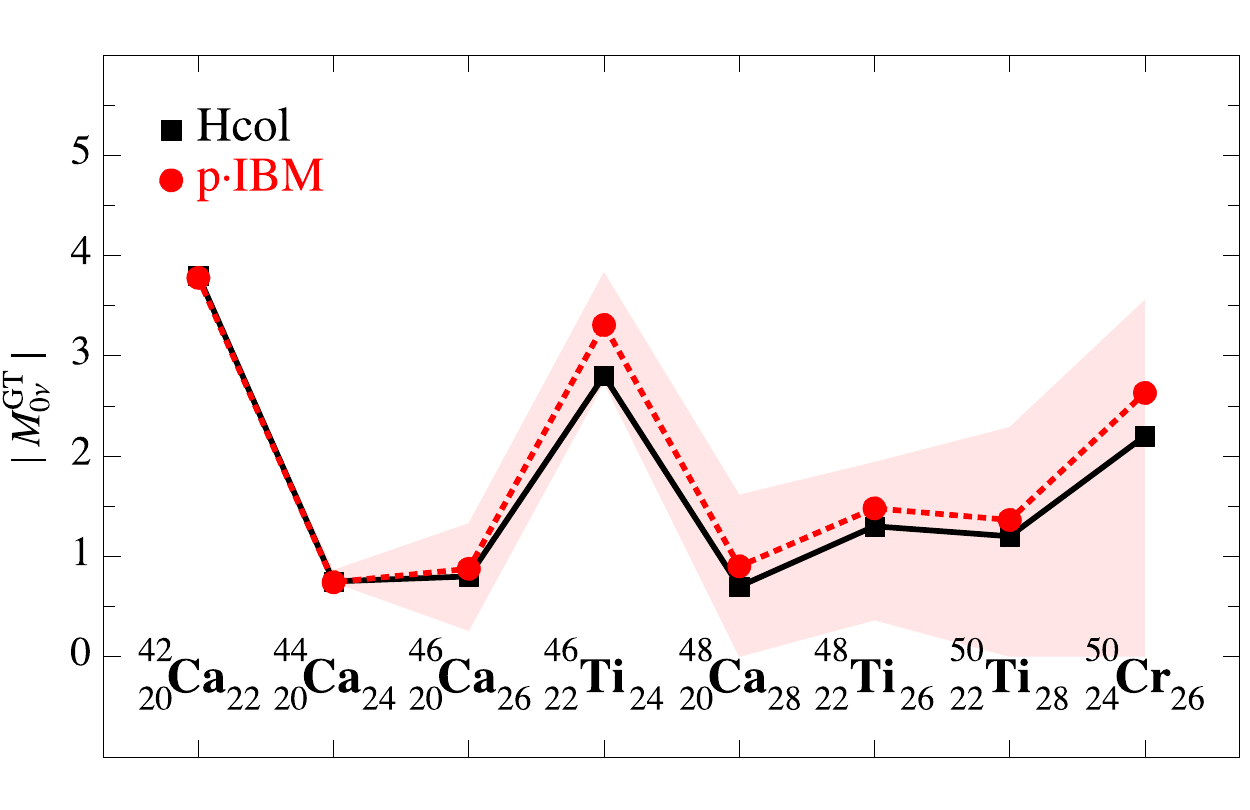}
\caption{Same as Fig.~\ref{f_b2}
for the collective Hamiltonian without (top) and with (bottom) isoscalar pairing,
and with IBM and \mbox{$p$-IBM} results appearing in the left and right panels, respectively.}
\label{f_coll}
\end{figure*}

\begin{figure}[b]
\centering
\includegraphics[width=\columnwidth]{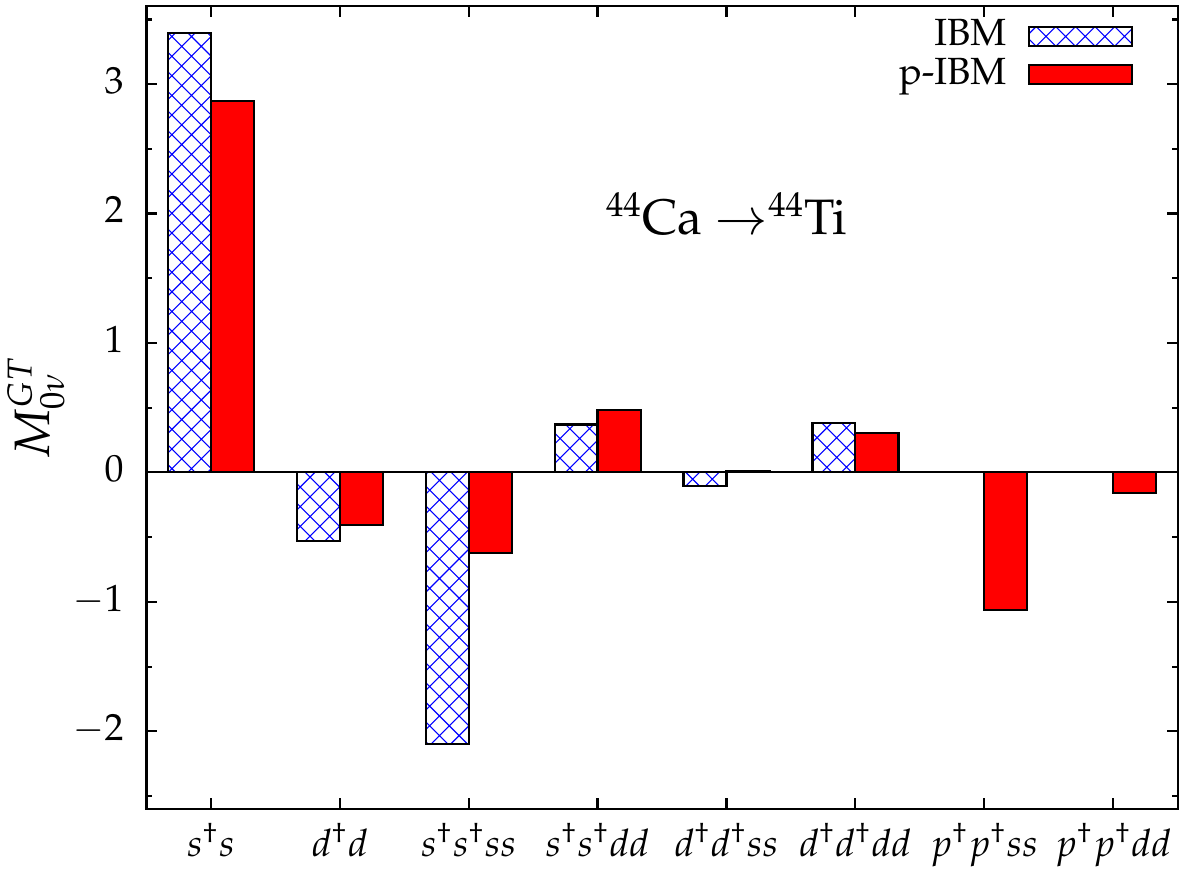}
\caption{\label{f_hist}Contributions to the Gamow-Teller matrix element
$M^{GT}_{0\nu}$ from different terms in the boson Hamiltonian (see text) for the
IBM (cross-hatched) and \mbox{$p$-IBM} (solid).}
\end{figure}  

The results for the total $0\nu\beta\beta$ matrix elements are shown in
Fig.~\ref{f_b2}.  Because isospin is conserved both in the shell model and in
the IBM, transition matrix elements for mirror sets of nuclei (e.g., $^{44}{\rm
Ca}\rightarrow{}^{44}{\rm Ti}$ and $^{44}{\rm Ti}\rightarrow{}^{44}{\rm Cr}$)
are equal and we show them for only one set here and in what follows.  The
shaded area in the figure indicates the values of the $0\nu\beta\beta$ matrix
elements obtained by varying the Hamiltonian and $0\nu\beta\beta$ operators
together between their bare and effective limits.  This area is very large in
the IBM and significantly reduced if effects of the $p$ boson are included.
Figure \ref{f_b2gt} shows the results for the Gamow-Teller part of the
$0\nu\beta\beta$ matrix elements, $M^{\rm GT}_{0\nu}$.  The \mbox{$p$-IBM} is
clearly superior to IBM in matching the shell-model trends, although it
systematically overestimates the $0\nu\beta\beta$ matrix elements.  One
conspicuous feature of the shell-model calculation is the enhancement of
transitions between mirror nuclei (i.e., $^{42}{\rm Ca}\rightarrow{}^{42}{\rm
Ti}$, $^{46}{\rm Ti}\rightarrow{}^{46}{\rm Cr}$, and $^{50}{\rm
Cr}\rightarrow{}^{50}{\rm Fe}$).  This \mbox{$p$-IBM} reproduces the resulting
``kink'' in the calculated set of matrix Gamow-Teller matrix elements, but the
IBM does not.  

Despite the better performance of the \mbox{$p$-IBM}, the range of possibilities
it predicts---reflected by the shaded areas that represent the plausible amount
of phenomenological modification to the mapped effective operators---is large
enough that one could question whether a $p$ boson is really essential in IBM
calculations of matrix elements for the more complicated nuclei that are used in
experiments.  To provide a better measure of the $p$-boson's importance, we
examine the degree to which the IBM captures the effects of isoscalar pairing
(shown repeatedly to be important for $\beta\beta$ decay~\cite{men16,vog86})
with and without the new degree of freedom.  There is not a unique prescription
for isolating the isoscalar-pairing part of the KB3G interaction, so we
substitute the multi-separable collective interaction employed in
Ref.~\cite{men16}.  This ``collective'' Hamiltonian supplements the KB3G
monopole part with separable like-particle pairing, isoscalar-pairing,
quadrupole-quadrupole , and spin-isospin interactions, with coefficients
determined through the methods presented in Ref.~\cite{duf96}.  With this
Hamiltonian, it is a simple matter to turn the isoscalar pairing on or off for
tests.

To carry out such tests, we repeat the entire mapping procedure with the new
Hamiltonian, with and without isoscalar pairing.  Figure~\ref{f_coll} shows the
results for the matrix elements of $M_{0\nu}^{\rm GT}$.  The left column
contrasts these matrix elements for our $sd$ IBM, with (bottom) and without
(top) isoscalar pairing.  Without isoscalar pairing the IBM reproduces the
shell-model matrix elements fairly well, and the range of predictions associated
with the shaded band, though not small, is not unreasonable.  When isoscalar
pairing is turned on, shell-model matrix elements shrink considerably, except
for the mirror transitions.  Though the IBM matrix elements also shrink on
average, the range of predictions is much larger.  

The results of the \mbox{$p$-IBM} in the right column are different, not so much in the
top figure, where, as expected, the $p$ boson makes little difference in the
absence of isoscalar pairing, but in the bottom.  When the $T=0$ pairing
interaction is on, the \mbox{$p$-IBM} with the best value of $x$ reproduces the shell
model results nearly perfectly, and the range of predictions is much smaller
than when the pairing is off or when the $p$ boson is absent.  Clearly, the $p$
boson is required to fully capture the effects of isoscalar pairing.  Even with
it, however, the range of predictions grows noticeably after the boson number
reaches about four.

We can look in a little more detail at how the two models attempt to capture the
physics of the Gamow-Teller matrix element.  Figure~\ref{f_hist} shows its
decomposition into pieces produced by different terms in the corresponding boson
operator for the decay of $^{44}$Ca.  The contribution of all terms involving
particular creation and destruction operators are grouped together, with all
possible values for the angular momentum and isospin of pairs of creation (and
annihilation) operators summed over.  As in the realistic \mbox{IBM-2}
calculations of Ref.~\cite{bar09}, the largest contribution is from the $s^\dag
s$ term, which converts a neutron $s$ boson into a proton $s$ boson.  And as in
those calculations, the contribution of $d^\dag d$, representing transitions
between $d$ bosons, is negative and relatively small.  Here, however, other
terms that are absent or suppressed in Ref.~\cite{bar09} contribute
significantly.  When the $p$ boson is included, the largest contribution,
outside of that from $s^\dag s$, is from $p^\dag p^\dag s s$.  This operator,
roughly speaking, replaces one neutron with a proton in each of two $J=0$
neutron pairs, while recoupling the resulting pairs to angular momentum $J=1$
and isospin $T=0$. The isoscalar-pairing interaction ensures that both the
initial and final configurations are well represented in the corresponding
ground states.  In the absence of a $p$ boson, the IBM attempts to mock up the
isoscalar pairs in the final nucleus by isovector proton-neutron $s$ bosons,
through the term $s^\dag s^\dag s s$.  Not surprisingly, the physics of
isoscalar pairing is not as well captured.  In the \mbox{IBM-2}, which does not
contain neutron-proton bosons of any kind, $d^\dag d$ is the only term
compensating $s^\dag s$.  One suspects that the effects of isoscalar pairing are
overlooked.

\section{Conclusions}
\label{s_conc}
Our results clearly suggest that although the isoscalar-pair bosons have a
deleterious effect on spectra---the inevitable result of diluting
collectivity---they are important for $\beta\beta$ decay.  In our calculations
with the realistic KB3G interaction, the improvement they offer is only modest,
but the reason, undoubtedly, is that our mapping is exact only for two and four
nucleons and we do not know how best to extrapolate it to larger numbers.  This
problem plagues almost all applications of the Lee Suzuki mapping procedure.  

What are the implications for the realistic \mbox{IBM-2} calculations of
Refs.~\cite{bar09,bar13,bar15,bar15a}?  Would they be improved by the addition
of a $p$ boson?  Isoscalar paring is probably more effective in light $pf$-shell
nuclei discussed here than, e.g., in $^{76}$Ge, so we have to be a little
careful in extrapolating blindly.  But many studies have shown isoscalar pairing
to play a role in all the nuclei used in experiments, and the $p$ boson thus has
the potential to improve the fidelity with which they are treated. 

A useful extension of the \mbox{IBM-2}, however, would require some careful
phenomenology.  The $0\nu\beta\beta$ operator in the realistic calculations,
like that here, comes from a mapping of states with a few valence nucleons, and
so should have similar properties to ours.  But the \mbox{IBM-2} Hamiltonian is
entirely phenomenological and without a guiding principle and careful fitting,
it is not obvious how best to modify it.  One might try to map the Hamiltonian
and select from the result the most important terms that contain $p$-boson
operators, and then modify the coefficients by fitting, e.g., to single-$\beta$
decay rates (which require their own mapped operator) or other observables.
Until an attempt is made, we cannot know how successful such a program would be.
Our results, however, imply that it would be worthwhile. 

\section*{Acknowledgments}
This work was partially supported (JE) by FUSTIPEN (French-U.S. Theory Institute
for Physics with Exotic Nuclei) under DOE grant No.\ DE-FG02-10ER41700.  JE also
acknowledges support form the US Department of Energy under Grant Nos.\
DE-FG02-97ER41019, DE-SC0008641, and DE-SC0004142.  KN acknowledges support from
JSPS and from the Marie Curie Actions grant within the Seventh Framework Program
of the European Commission under Grant No.\  PIEF-GA-2012-327398.

\appendix
\section{Matrix elements in a two-pair basis}
\label{a_mes}
We summarize in this appendix the expressions for the matrix element of a
generic one- or two-body operator between two-pair states.  For $n=2$ we rewrite
the pair state~(\ref{e_bas1}) in a more explicitly as 
\begin{align}
&\ket{\gamma_a\gamma_b\gamma_c\gamma_d[\Gamma_1\Gamma_2]\Lambda M_\Lambda}\equiv
\ket{abcd[\Gamma_1\Gamma_2]\Lambda M_\Lambda}
\label{e_abas1}\\&
\propto
{\cal A}\sum_{M_1M_2}
(\Gamma_1M_1\,\Gamma_2M_2|\Lambda M_\Lambda)
\ket{ab;\Gamma_1M_1}
\ket{cd;\Gamma_2M_2},
\nonumber
\end{align}
where the pair states on the last line are normalized and anti-symmetric, and ${\cal A}$
is a four-nucleon anti-symmetrization operator.  Coupling to definite
angular momentum and isospin together with anti-symmetrization leads to an
expansion in terms of coefficients of fractional parentage (CFPs),
\begin{align}
&\ket{abcd[\Gamma_1\Gamma_2]\Lambda M_\Lambda}
\nonumber\\&\qquad=
\sum_{qrst}\sum_{\bar\Gamma'_1\bar\Gamma'_2}
[qr(\bar\Gamma'_1)st(\bar\Gamma'_2)\Lambda|\}abcd[\Gamma_1\Gamma_2]\Lambda]
\nonumber\\&\qquad\phantom{=}\times
\ket{qr(\bar\Gamma'_1)st(\bar\Gamma'_2);\Lambda M_\Lambda},
\label{e_abas2}
\end{align}
where the sum $\{qrst\}$ is over all permutations of $\{abcd\}$.  

Consider now an operator $\hat T^{(\lambda)}_{m_\lambda}$, where $\lambda$
refers to the operator's tensor character in angular momentum and isospin, and
$m_\lambda$ to the respective projections.  By virtue of the Wigner-Eckart
theorem~\cite{talmi93}, the matrix elements of $\hat T^{(\lambda)}_{m_\lambda}$
can be written as 
\begin{align}
&\braket{a'b'c'd'[\Gamma'_1\Gamma'_2]\Lambda'M'_\Lambda|
\hat T^{(\lambda)}_{m_\lambda}
|a''b''c''d''[\Gamma''_1\Gamma''_2]\Lambda''M''_\Lambda}
\nonumber\\&=
(-)^{\Lambda'-M'_\Lambda}
\Bigl(\begin{array}{ccc}\Lambda'&\lambda&\Lambda''\\-M'_\Lambda&m_\lambda&M''_\Lambda\end{array}\Bigr)
\nonumber\\&\phantom{=}\times
\braket{a'b'c'd'[\Gamma'_1\Gamma'_2]\Lambda'\| \hat T^{(\lambda)}
\|a''b''c''d''[\Gamma''_1\Gamma''_2]\Lambda''}.
\label{e_amat1}
\end{align}
The matrix element on the right side of this equation is reduced in angular
momentum $J$ and isospin $T$ and the symbol in round brackets, consequently,
refers to a product of Wigner coefficients, one pertaining to $J$ and the other
to $T$.  The reduced matrix element in turn can be expressed as
\begin{align}
&\braket{a'b'c'd'[\Gamma'_1\Gamma'_2]\Lambda'\|
\hat T^{(\lambda)}
\|a''b''c''d''[\Gamma''_1\Gamma''_2]\Lambda''}
\nonumber\\&=
p\sum_{q'r's't'}\sum_{\bar\Gamma'_1\bar\Gamma'_2}
[q'r'(\bar\Gamma'_1)s't'(\bar\Gamma'_2)\Lambda'|\}a'b'c'd'[\Gamma'_1\Gamma'_2]\Lambda']
\nonumber\\&\phantom{=}\times
\sum_{q''r''s''t''}\sum_{\bar\Gamma''_2}
[q''r''(\bar\Gamma'_1)s''t''(\bar\Gamma''_2)\Lambda''|\}a''b''c''d''[\Gamma''_1\Gamma''_2]\Lambda'']
\nonumber\\&\phantom{=}\times
\braket{ q'r';\bar\Gamma'_1|q''r'';\bar\Gamma'_1}
\braket{ s't';\bar\Gamma'_2\|\hat T^{(\lambda)}\|s''t'';\bar\Gamma''_2}
\nonumber\\&\phantom{=}\times
(-)^{\bar\Gamma'_1+\bar\Gamma''_2+\Lambda'+\lambda}[\Lambda'][\Lambda'']
\Bigl\{\begin{array}{ccc}\bar\Gamma'_2&\Lambda'&\bar\Gamma'_1\\\Lambda''&\bar\Gamma''_2&\lambda\end{array}\Bigr\}\,,
\label{e_amat2}
\end{align}
with $[x]\equiv\sqrt{2x+1}$, $p=2$ for a one-body and $p=6$ for a two-body
operator, and 
\begin{align}
&\braket{ q'r';\bar\Gamma'_1|q''r'';\bar\Gamma'_1}
\label{e_ad}\\&=
\frac{1}{1+\delta_{q'r'}}
\left(\delta_{q'q''}\delta_{r'r''}-(-)^{\gamma_{q'}+\gamma_{r'}-\bar\Gamma'_1}\delta_{q'r''}\delta_{q'r''}\right).
\nonumber
\end{align}
The symbol in curly brackets in Eq.~(\ref{e_amat2}) refers to a product of Racah
coefficients~\cite{talmi93} in angular momentum and isospin space,
\begin{equation}
\Bigl\{\begin{array}{ccc}\Gamma'_2&\Lambda'&\Gamma'_1\\\Lambda''&\Gamma''_2&\lambda\end{array}\Bigr\}\equiv
\Bigl\{\begin{array}{ccc}J'_2&J'&J'_1\\J''&J''_2&\lambda_j\end{array}\Bigr\}
\Bigl\{\begin{array}{ccc}T'_2&T'&T'_1\\T''&T''_2&\lambda_t\end{array}\Bigr\}\,.
\label{e_racah}
\end{equation}
An important case occurs if $(\lambda)=(0,0)$, that is, the tensor operator is
scalar in angular momentum as well as isospin.  Then
$\Lambda'=\Lambda''\equiv\Lambda$ and the expression~(\ref{e_amat2}) for the
matrix element reduces to
\begin{align}
&\braket{a'b'c'd'[\Gamma'_1\Gamma'_2]\Lambda|
\hat T^{(0)}_0
|a''b''c''d''[\Gamma''_1\Gamma''_2]\Lambda}
\nonumber\\&=
p\sum_{q'r's't'}\sum_{\bar\Gamma'_1\bar\Gamma'_2}
[q'r'(\bar\Gamma'_1)s't'(\bar\Gamma'_2)\Lambda|\}a'b'c'd'[\Gamma'_1\Gamma'_2]\Lambda]
\nonumber\\&\phantom{=}\times
\sum_{q''r''s''t''}
[q''r''(\bar\Gamma'_1)s''t''(\bar\Gamma'_2)\Lambda|\}a''b''c''d''[\Gamma''_1\Gamma''_2]\Lambda]
\nonumber\\&\phantom{=}\times
\braket{ q'r';\bar\Gamma'_1|q''r'';\bar\Gamma'_1}
\braket{ s't';\bar\Gamma'_2|\hat T^{(0)}_0|s''t'';\bar\Gamma'_2}.
\label{e_amat3}
\end{align}
This formula (with $p=6$) applies to the matrix elements of a scalar two-body
interaction, in which case the last factor in Eq.~(\ref{e_amat3}) is the
two-body matrix element,
\begin{equation}
\braket{ s't';\Gamma|\hat T^{(0)}_0|s''t'';\Gamma}=
\upsilon^\Gamma_{s't's''t''}\,.
\label{e_amat4}
\end{equation}
The $0\nu\beta\beta$ operator can be assumed scalar in angular momentum but not in
isospin, and therefore requires the application of the more general
expression in Eq.\ (\ref{e_amat2}).

\section{Order-by-order mapping of non-scalar operators}
\label{a_order}
For a non-scalar operator,it is better to define the boson image by requiring
the equality of {\em reduced} matrix elements in angular momentum and isospin,
defined through the Wigner-Eckart theorem~\cite{talmi93}.  The one-boson term
follows from
\begin{equation}
\braket{ b_{\Gamma'}\|\hat T^{{\rm b}(\lambda)}\| b_{\Gamma''}}=
\braket{\bar b_{\Gamma'}\|\hat T^{{\rm b}(\lambda)}\|\bar b_{\Gamma''}}\doteq
\braket{\bar B_{\Gamma'}\|\hat T^{{\rm f}(\lambda)}\|\bar B_{\Gamma''}}.
\label{e_ord1}
\end{equation}
The fermion matrix element on the right-hand side of Eq.~(\ref{e_ord1}) is given
by
\begin{align}
&\braket{\bar B_{\Gamma'}\|\hat T^{{\rm f}(\lambda)}\|\bar B_{\Gamma''}}
\nonumber\\&=
\sum_{\gamma'_1\gamma'_2}\sum_{\gamma''_1\gamma''_2}
\bar\alpha^{\Gamma'}_{\gamma'_1\gamma'_2}
\bar\alpha^{\Gamma''}_{\gamma''_1\gamma''_2}
(-)^{\Gamma'-M'_\Gamma}
\Bigl(\begin{array}{ccc}\Gamma'&\lambda&\Gamma''\\-M'_\Gamma&m_\lambda&M''_\Gamma\end{array}\Bigr)^{-1}
\nonumber\\&\phantom{=}\times
\braket{\gamma'_1\gamma'_2;\Gamma'M'_\Gamma|\hat T^{{\rm
f}(\lambda)}_{m_\lambda}|\gamma''_1\gamma''_2;\Gamma''M''_\Gamma},
\label{e_ord2}
\end{align}
where $\braket{\gamma'_1\gamma'_2;\Gamma'M'_\Gamma| \hat T^{{\rm
f}(\lambda)}_{m_\lambda}|\gamma''_1\gamma''_2;\Gamma''M''_\Gamma}$ are matrix
elements in the complete shell-model space and
$\bar\alpha^\Gamma_{\gamma_1\gamma_2}$ are structure coefficients of normalized
collective pairs,
\begin{equation}
\bar B_{\Gamma M_\Gamma}^\dag\ket{{\rm O}}=
\sum_{\gamma_1\gamma_2}
\bar\alpha^\Gamma_{\gamma_1\gamma_2}
\ket{\gamma_1\gamma_2;\Gamma M_\Gamma}.
\label{e_ord3}
\end{equation}
Equations~(\ref{e_ord1}) and~(\ref{e_ord2}) define entirely the one-body part
$\hat T^{{\rm b}(\lambda)}_{1,m_\lambda}$ of the mapped boson operator.  This
object can be written in second quantization as
\begin{equation}
\hat T^{{\rm b}(\lambda)}_{1,m_\lambda}=
\sum_{\Gamma'\Gamma''}t_{\Gamma'\Gamma''}
(b_{\Gamma'}^\dag\times\tilde b_{\Gamma''})^{(\lambda)}_{m_\lambda}\,,
\label{e_ord4}
\end{equation}
with
\begin{equation}
t_{\Gamma'\Gamma''}\equiv
\frac{\braket{\bar B_{\Gamma'}\|\hat T^{{\rm f}(\lambda)}\|\bar
B_{\Gamma''}}}{\sqrt{2\lambda+1}},
\label{e_ord5}
\end{equation}
an expression showing that in general $\hat T^{{\rm b}(\lambda)}_{1,m_\lambda}$
is non-diagonal in the boson basis.

The two-body part of the mapped boson operator follows from the obvious identity
\begin{align}
&\braket{ b_{\Gamma'_1}b_{\Gamma'_2};\Lambda'\|\hat T^{{\rm b}(\lambda)}\|b_{\Gamma''_1}b_{\Gamma''_2};\Lambda''}
\nonumber\\&\quad=
\braket{b_{\Gamma'_1}b_{\Gamma'_2};\Lambda'\|
\hat T^{{\rm b}(\lambda)}_1+\hat T^{{\rm
b}(\lambda)}_2\|b_{\Gamma''_1}b_{\Gamma''_2};\Lambda''}.
\label{e_ord6}
\end{align}
The matrix element on the left side is the boson image of the fermion operator
and can be computed from Eq.~(\ref{e_map1}).  By using the operator
representation in Eq.\ (\ref{e_ord4}), one can work out the first (one-body)
term on the right side, obtaining
\begin{align}
&\braket{ b_{\Gamma'_1}b_{\Gamma'_2};\Lambda'\|\hat T^{{\rm b}(\lambda)}_1\|b_{\Gamma''_1}b_{\Gamma''_2};\Lambda''}
\label{e_ord7}\\&=
[\Lambda'][\lambda][\Lambda'']
\hat P(-)^{\Gamma'_1+\Gamma'_2+\Lambda''+\lambda}
t_{\Gamma'_1\Gamma''_1}
\Bigl\{\begin{array}{ccc}\Gamma'_1&\Lambda'&\Gamma'_2\\\Lambda''&\Gamma''_1&\lambda\end{array}\Bigr\}
\delta_{\Gamma'_2\Gamma''_2}\,,
\nonumber
\end{align}
where the operator $\hat P$ takes care of the different permutations: $\hat
P\equiv\hat P_{\Gamma'_1\Gamma'_2\Lambda'}\hat
P_{\Gamma''_1\Gamma''_2\Lambda''}$ with
 \begin{equation}
\hat P_{\Gamma_1\Gamma_2\Lambda}\equiv
\frac{f(\Gamma_1,\Gamma_2,\Lambda)+(-)^{\Gamma_1+\Gamma_2-\Lambda}f(\Gamma_2,\Gamma_1,\Lambda)}
{\sqrt{1+\delta_{\Gamma_1\Gamma_2}}}.
\label{e_ord8}
\end{equation}
Equation~(\ref{e_ord6}) therefore entirely defines the two-body part
$\hat T^{{\rm b}(\lambda)}_{2,m_\lambda}$ of the mapped boson operator.

As an example, we apply the above formulas to the $0\nu\beta\beta$ operator, which is
a non-scalar tensor $\hat T^{(\lambda)}_{m_\lambda}$ with $\lambda=(0,2)$ and
$m_\lambda=(0,-2)$.  It is of two-body character in the fermions and we
calculate its image up to two-body terms in the bosons.  We assume, as is the
case in the applications discussed in this paper, that off-diagonal matrix
elements between pair states vanish, that is, that $\braket{\bar
B_{\Gamma'}\|\hat T^{{\rm f}\beta\beta}\|\bar B_{\Gamma''}}=0$ if
$\Gamma'\neq\Gamma''$.  This relation obtains because the pairs have different
angular momenta ($S$, $D$ and $P$) and the $0\nu\beta\beta$ operator is assumed scalar
in angular momentum.  For two-particle states, $0\nu\beta\beta$ decay takes place from an
initial state with $T=1,M''_T=+1$ to a final state with $T=1,M'_T=-1$, and the
matrix element~(\ref{e_ord2}) reduces to
\begin{align}
&\braket{\bar B_{JT}\|\hat T^{{\rm f}\beta\beta}\|\bar B_{JT}}
\label{e_ord9}\\&\quad=
\sqrt{5(2J+1)}\braket{ JT,M'_T=-1|\hat T^{{\rm f}\beta\beta}|JT,M''_T=+1}.
\nonumber
\end{align}
The contribution~(\ref{e_ord7}) of the one-body part of the boson operator
between two-boson states also simplifies because
$\Gamma'_1=\Gamma''_1\equiv\Gamma_1$ and $\Gamma'_2=\Gamma''_2\equiv\Gamma_2$,
and can be written explicitly as
\begin{widetext}
\begin{equation}
\braket{ b_{\Gamma_1}b_{\Gamma_2};\Lambda'\|\hat T^{{\rm b}(\lambda)}_1\|b_{\Gamma_1}b_{\Gamma_2};\Lambda''}=
[\Lambda'][\lambda][\Lambda''](-)^{\Gamma_1+\Gamma_2+\lambda}
\Bigl[(-)^{\Lambda''}t_{\Gamma_1\Gamma_1}
\Bigl\{\begin{array}{ccc}\Gamma_1&\Lambda'&\Gamma_2\\\Lambda''&\Gamma_1&\lambda\end{array}\Bigr\}+
(-)^{\Lambda'}t_{\Gamma_2\Gamma_2}
\Bigl\{\begin{array}{ccc}\Gamma_2&\Lambda'&\Gamma_1\\\Lambda''&\Gamma_2&\lambda\end{array}\Bigr\}\Bigr]\,.
\label{e_ord10}
\end{equation}
\end{widetext}


\begin{thebibliography}{42}%
\makeatletter
\providecommand \@ifxundefined [1]{%
 \@ifx{#1\undefined}
}%
\providecommand \@ifnum [1]{%
 \ifnum #1\expandafter \@firstoftwo
 \else \expandafter \@secondoftwo
 \fi
}%
\providecommand \@ifx [1]{%
 \ifx #1\expandafter \@firstoftwo
 \else \expandafter \@secondoftwo
 \fi
}%
\providecommand \natexlab [1]{#1}%
\providecommand \enquote  [1]{``#1''}%
\providecommand \bibnamefont  [1]{#1}%
\providecommand \bibfnamefont [1]{#1}%
\providecommand \citenamefont [1]{#1}%
\providecommand \href@noop [0]{\@secondoftwo}%
\providecommand \href [0]{\begingroup \@sanitize@url \@href}%
\providecommand \@href[1]{\@@startlink{#1}\@@href}%
\providecommand \@@href[1]{\endgroup#1\@@endlink}%
\providecommand \@sanitize@url [0]{\catcode `\\12\catcode `\$12\catcode
  `\&12\catcode `\#12\catcode `\^12\catcode `\_12\catcode `\%12\relax}%
\providecommand \@@startlink[1]{}%
\providecommand \@@endlink[0]{}%
\providecommand \url  [0]{\begingroup\@sanitize@url \@url }%
\providecommand \@url [1]{\endgroup\@href {#1}{\urlprefix }}%
\providecommand \urlprefix  [0]{URL }%
\providecommand \Eprint [0]{\href }%
\providecommand \doibase [0]{http://dx.doi.org/}%
\providecommand \selectlanguage [0]{\@gobble}%
\providecommand \bibinfo  [0]{\@secondoftwo}%
\providecommand \bibfield  [0]{\@secondoftwo}%
\providecommand \translation [1]{[#1]}%
\providecommand \BibitemOpen [0]{}%
\providecommand \bibitemStop [0]{}%
\providecommand \bibitemNoStop [0]{.\EOS\space}%
\providecommand \EOS [0]{\spacefactor3000\relax}%
\providecommand \BibitemShut  [1]{\csname bibitem#1\endcsname}%
\let\auto@bib@innerbib\@empty
\bibitem [{\citenamefont {Henning}(2016)}]{Henning16}%
  \BibitemOpen
  \bibfield  {author} {\bibinfo {author} {\bibfnamefont {R.}~\bibnamefont
  {Henning}},\ }\href {\doibase 10.1016/j.revip.2016.03.001} {\bibfield
  {journal} {\bibinfo  {journal} {Rev. Phys.}\ }\textbf {\bibinfo {volume}
  {1}},\ \bibinfo {pages} {29} (\bibinfo {year} {2016})}\BibitemShut {NoStop}%
\bibitem [{\citenamefont {Dell'Oro}\ \emph {et~al.}(2016)\citenamefont
  {Dell'Oro}, \citenamefont {Marcocci}, \citenamefont {Viel},\ and\
  \citenamefont {Vissani}}]{DellOro16}%
  \BibitemOpen
  \bibfield  {author} {\bibinfo {author} {\bibfnamefont {S.}~\bibnamefont
  {Dell'Oro}}, \bibinfo {author} {\bibfnamefont {S.}~\bibnamefont {Marcocci}},
  \bibinfo {author} {\bibfnamefont {M.}~\bibnamefont {Viel}}, \ and\ \bibinfo
  {author} {\bibfnamefont {F.}~\bibnamefont {Vissani}},\ }\href {\doibase
  10.1155/2016/2162659} {\bibfield  {journal} {\bibinfo  {journal} {Adv. High
  Energy Phys.}\ }\textbf {\bibinfo {volume} {2016}},\ \bibinfo {pages}
  {2162659} (\bibinfo {year} {2016})}\BibitemShut {NoStop}%
\bibitem [{\citenamefont {Cremonesi}\ and\ \citenamefont
  {Pavan}(2014)}]{Cremonesi14}%
  \BibitemOpen
  \bibfield  {author} {\bibinfo {author} {\bibfnamefont {O.}~\bibnamefont
  {Cremonesi}}\ and\ \bibinfo {author} {\bibfnamefont {M.}~\bibnamefont
  {Pavan}},\ }\href {\doibase 10.1155/2014/951432} {\bibfield  {journal}
  {\bibinfo  {journal} {Adv. High Energy Phys.}\ }\textbf {\bibinfo {volume}
  {2014}},\ \bibinfo {pages} {951432} (\bibinfo {year} {2014})}\BibitemShut
  {NoStop}%
\bibitem [{\citenamefont {G{\'o}mez-Cadenas}\ \emph {et~al.}(2012)\citenamefont
  {G{\'o}mez-Cadenas}, \citenamefont {Mart{\'i}n-Albo}, \citenamefont
  {Mezzetto}, \citenamefont {Monrabal},\ and\ \citenamefont
  {Sorel}}]{GomezCadenas12}%
  \BibitemOpen
  \bibfield  {author} {\bibinfo {author} {\bibfnamefont {J.~J.}\ \bibnamefont
  {G{\'o}mez-Cadenas}}, \bibinfo {author} {\bibfnamefont {J.}~\bibnamefont
  {Mart{\'i}n-Albo}}, \bibinfo {author} {\bibfnamefont {M.}~\bibnamefont
  {Mezzetto}}, \bibinfo {author} {\bibfnamefont {F.}~\bibnamefont {Monrabal}},
  \ and\ \bibinfo {author} {\bibfnamefont {M.}~\bibnamefont {Sorel}},\ }\href
  {\doibase 10.1393/ncr/i2012-10074-9} {\bibfield  {journal} {\bibinfo
  {journal} {Riv. Nuovo Cim.}\ }\textbf {\bibinfo {volume} {35}},\ \bibinfo
  {pages} {29} (\bibinfo {year} {2012})}\BibitemShut {NoStop}%
\bibitem [{\citenamefont {Schechter}\ and\ \citenamefont
  {Valle}(1982)}]{sch82}%
  \BibitemOpen
  \bibfield  {author} {\bibinfo {author} {\bibfnamefont {J.}~\bibnamefont
  {Schechter}}\ and\ \bibinfo {author} {\bibfnamefont {J.~W.~F.}\ \bibnamefont
  {Valle}},\ }\href {\doibase 10.1103/PhysRevD.25.774} {\bibfield  {journal}
  {\bibinfo  {journal} {Phys.\ Rev. D}\ }\textbf {\bibinfo {volume} {25}},\
  \bibinfo {pages} {2951} (\bibinfo {year} {1982})}\BibitemShut {NoStop}%
\bibitem [{\citenamefont {Bilenky}\ and\ \citenamefont {Petcov}(1987)}]{bil87}%
  \BibitemOpen
  \bibfield  {author} {\bibinfo {author} {\bibfnamefont {S.~M.}\ \bibnamefont
  {Bilenky}}\ and\ \bibinfo {author} {\bibfnamefont {S.~T.}\ \bibnamefont
  {Petcov}},\ }\href@noop {} {\bibfield  {journal} {\bibinfo  {journal} {Rev.
  Mod. Phys.}\ }\textbf {\bibinfo {volume} {59}},\ \bibinfo {pages} {671}
  (\bibinfo {year} {1987})}\BibitemShut {NoStop}%
\bibitem [{\citenamefont {{Avignone{ }III}}\ \emph {et~al.}(2008)\citenamefont
  {{Avignone{ }III}}, \citenamefont {Elliott},\ and\ \citenamefont
  {Engel}}]{avi08}%
  \BibitemOpen
  \bibfield  {author} {\bibinfo {author} {\bibfnamefont {F.~T.}\ \bibnamefont
  {{Avignone{ }III}}}, \bibinfo {author} {\bibfnamefont {S.~R.}\ \bibnamefont
  {Elliott}}, \ and\ \bibinfo {author} {\bibfnamefont {J.}~\bibnamefont
  {Engel}},\ }\href {\doibase 10.1103/RevModPhys.80.481} {\bibfield  {journal}
  {\bibinfo  {journal} {Rev.\ Mod.\ Phys.}\ }\textbf {\bibinfo {volume} {80}},\
  \bibinfo {pages} {481} (\bibinfo {year} {2008})}\BibitemShut {NoStop}%
\bibitem [{\citenamefont {Deppisch}\ and\ \citenamefont
  {Suhonen}(2016)}]{Deppisch16}%
  \BibitemOpen
  \bibfield  {author} {\bibinfo {author} {\bibfnamefont {F.~F.}\ \bibnamefont
  {Deppisch}}\ and\ \bibinfo {author} {\bibfnamefont {J.}~\bibnamefont
  {Suhonen}},\ }\href {\doibase 10.1103/PhysRevC.94.055501} {\bibfield
  {journal} {\bibinfo  {journal} {Phys. Rev. C}\ }\textbf {\bibinfo {volume}
  {94}},\ \bibinfo {pages} {055501} (\bibinfo {year} {2016})}\BibitemShut
  {NoStop}%
\bibitem [{\citenamefont {Engel}\ and\ \citenamefont
  {Men\'endez}(2017)}]{engel16}%
  \BibitemOpen
  \bibfield  {author} {\bibinfo {author} {\bibfnamefont {J.}~\bibnamefont
  {Engel}}\ and\ \bibinfo {author} {\bibfnamefont {J.}~\bibnamefont
  {Men\'endez}},\ }\href {\doibase 10.1088/1361-6633/aa5bc5} {\bibfield
  {journal} {\bibinfo  {journal} {Rept. Prog. Phys.}\ }\textbf {\bibinfo
  {volume} {80}},\ \bibinfo {pages} {046301} (\bibinfo {year} {2017})},\
  \Eprint {http://arxiv.org/abs/1610.06548} {arXiv:1610.06548 [nucl-th]}
  \BibitemShut {NoStop}%
\bibitem [{\citenamefont {Iachello}\ and\ \citenamefont
  {Arima}(1987)}]{Iachello87}%
  \BibitemOpen
  \bibfield  {author} {\bibinfo {author} {\bibfnamefont {F.}~\bibnamefont
  {Iachello}}\ and\ \bibinfo {author} {\bibfnamefont {A.}~\bibnamefont
  {Arima}},\ }\href@noop {} {\emph {\bibinfo {title} {The interacting boson
  model}}}\ (\bibinfo  {publisher} {Cambridge University Press},\ \bibinfo
  {address} {Cambridge},\ \bibinfo {year} {1987})\BibitemShut {NoStop}%
\bibitem [{\citenamefont {Barea}\ and\ \citenamefont {Iachello}(2009)}]{bar09}%
  \BibitemOpen
  \bibfield  {author} {\bibinfo {author} {\bibfnamefont {J.}~\bibnamefont
  {Barea}}\ and\ \bibinfo {author} {\bibfnamefont {F.}~\bibnamefont
  {Iachello}},\ }\href {\doibase 10.1103/PhysRevC.79.044301} {\bibfield
  {journal} {\bibinfo  {journal} {Phys.\ Rev.\ C}\ }\textbf {\bibinfo {volume}
  {79}},\ \bibinfo {pages} {044301} (\bibinfo {year} {2009})}\BibitemShut
  {NoStop}%
\bibitem [{\citenamefont {Barea}\ \emph {et~al.}(2013)\citenamefont {Barea},
  \citenamefont {Kotila},\ and\ \citenamefont {Iachello}}]{bar13}%
  \BibitemOpen
  \bibfield  {author} {\bibinfo {author} {\bibfnamefont {J.}~\bibnamefont
  {Barea}}, \bibinfo {author} {\bibfnamefont {J.}~\bibnamefont {Kotila}}, \
  and\ \bibinfo {author} {\bibfnamefont {F.}~\bibnamefont {Iachello}},\ }\href
  {\doibase 10.1103/PhysRevC.87.014315} {\bibfield  {journal} {\bibinfo
  {journal} {Phys.\ Rev.\ C}\ }\textbf {\bibinfo {volume} {87}},\ \bibinfo
  {pages} {014315} (\bibinfo {year} {2013})}\BibitemShut {NoStop}%
\bibitem [{\citenamefont {Barea}\ \emph
  {et~al.}(2015{\natexlab{a}})\citenamefont {Barea}, \citenamefont {Kotila},\
  and\ \citenamefont {Iachello}}]{bar15}%
  \BibitemOpen
  \bibfield  {author} {\bibinfo {author} {\bibfnamefont {J.}~\bibnamefont
  {Barea}}, \bibinfo {author} {\bibfnamefont {J.}~\bibnamefont {Kotila}}, \
  and\ \bibinfo {author} {\bibfnamefont {F.}~\bibnamefont {Iachello}},\ }\href
  {\doibase 10.1103/PhysRevC.91.034304} {\bibfield  {journal} {\bibinfo
  {journal} {Phys. Rev. C}\ }\textbf {\bibinfo {volume} {91}},\ \bibinfo
  {pages} {034304} (\bibinfo {year} {2015}{\natexlab{a}})}\BibitemShut
  {NoStop}%
\bibitem [{\citenamefont {Arima}\ \emph {et~al.}(1977)\citenamefont {Arima},
  \citenamefont {Ohtsuka}, \citenamefont {Iachello},\ and\ \citenamefont
  {Talmi}}]{Arima77}%
  \BibitemOpen
  \bibfield  {author} {\bibinfo {author} {\bibfnamefont {A.}~\bibnamefont
  {Arima}}, \bibinfo {author} {\bibfnamefont {T.}~\bibnamefont {Ohtsuka}},
  \bibinfo {author} {\bibfnamefont {F.}~\bibnamefont {Iachello}}, \ and\
  \bibinfo {author} {\bibfnamefont {I.}~\bibnamefont {Talmi}},\ }\href
  {\doibase 10.1016/0370-2693(77)90860-7} {\bibfield  {journal} {\bibinfo
  {journal} {Phys. Lett.}\ }\textbf {\bibinfo {volume} {66B}},\ \bibinfo
  {pages} {205} (\bibinfo {year} {1977})}\BibitemShut {NoStop}%
\bibitem [{\citenamefont {Dellagiacoma}\ and\ \citenamefont
  {Iachello}(1989)}]{dellagiacoma89}%
  \BibitemOpen
  \bibfield  {author} {\bibinfo {author} {\bibfnamefont {F.}~\bibnamefont
  {Dellagiacoma}}\ and\ \bibinfo {author} {\bibfnamefont {F.}~\bibnamefont
  {Iachello}},\ }\href {\doibase
  http://dx.doi.org/10.1016/0370-2693(89)91434-2} {\bibfield  {journal}
  {\bibinfo  {journal} {Physics Letters B}\ }\textbf {\bibinfo {volume}
  {218}},\ \bibinfo {pages} {399 } (\bibinfo {year} {1989})}\BibitemShut
  {NoStop}%
\bibitem [{\citenamefont {Vogel}\ and\ \citenamefont
  {Zirnbauer}(1986)}]{vog86}%
  \BibitemOpen
  \bibfield  {author} {\bibinfo {author} {\bibfnamefont {P.}~\bibnamefont
  {Vogel}}\ and\ \bibinfo {author} {\bibfnamefont {M.~R.}\ \bibnamefont
  {Zirnbauer}},\ }\href {\doibase 10.1103/PhysRevLett.57.3148} {\bibfield
  {journal} {\bibinfo  {journal} {Phys. Rev. Lett.}\ }\textbf {\bibinfo
  {volume} {57}},\ \bibinfo {pages} {3148} (\bibinfo {year}
  {1986})}\BibitemShut {NoStop}%
\bibitem [{\citenamefont {Engel}\ \emph {et~al.}(1988)\citenamefont {Engel},
  \citenamefont {Vogel},\ and\ \citenamefont {Zirnbauer}}]{eng88}%
  \BibitemOpen
  \bibfield  {author} {\bibinfo {author} {\bibfnamefont {J.}~\bibnamefont
  {Engel}}, \bibinfo {author} {\bibfnamefont {P.}~\bibnamefont {Vogel}}, \ and\
  \bibinfo {author} {\bibfnamefont {M.~R.}\ \bibnamefont {Zirnbauer}},\ }\href
  {\doibase 10.1103/PhysRevC.37.731} {\bibfield  {journal} {\bibinfo  {journal}
  {Phys.\ Rev. C}\ }\textbf {\bibinfo {volume} {37}},\ \bibinfo {pages} {731}
  (\bibinfo {year} {1988})}\BibitemShut {NoStop}%
\bibitem [{\citenamefont {Hinohara}\ and\ \citenamefont {Engel}(2014)}]{hin14}%
  \BibitemOpen
  \bibfield  {author} {\bibinfo {author} {\bibfnamefont {N.}~\bibnamefont
  {Hinohara}}\ and\ \bibinfo {author} {\bibfnamefont {J.}~\bibnamefont
  {Engel}},\ }\href {\doibase 10.1103/PhysRevC.90.031301} {\bibfield  {journal}
  {\bibinfo  {journal} {Phys. Rev. C}\ }\textbf {\bibinfo {volume} {90}},\
  \bibinfo {pages} {031301} (\bibinfo {year} {2014})}\BibitemShut {NoStop}%
\bibitem [{\citenamefont {Men{\'e}ndez}\ \emph {et~al.}(2016)\citenamefont
  {Men{\'e}ndez}, \citenamefont {Hinohara}, \citenamefont {Engel},
  \citenamefont {Mart{\'i}nez-Pinedo},\ and\ \citenamefont
  {Rodr{\'i}guez}}]{men16}%
  \BibitemOpen
  \bibfield  {author} {\bibinfo {author} {\bibfnamefont {J.}~\bibnamefont
  {Men{\'e}ndez}}, \bibinfo {author} {\bibfnamefont {N.}~\bibnamefont
  {Hinohara}}, \bibinfo {author} {\bibfnamefont {J.}~\bibnamefont {Engel}},
  \bibinfo {author} {\bibfnamefont {G.}~\bibnamefont {Mart{\'i}nez-Pinedo}}, \
  and\ \bibinfo {author} {\bibfnamefont {T.~R.}\ \bibnamefont
  {Rodr{\'i}guez}},\ }\href {\doibase 10.1103/PhysRevC.93.014305} {\bibfield
  {journal} {\bibinfo  {journal} {Phys. Rev. C}\ }\textbf {\bibinfo {volume}
  {93}},\ \bibinfo {pages} {014305} (\bibinfo {year} {2016})}\BibitemShut
  {NoStop}%
\bibitem [{\citenamefont {Chen}(1993)}]{Chen93}%
  \BibitemOpen
  \bibfield  {author} {\bibinfo {author} {\bibfnamefont {J.-Q.}\ \bibnamefont
  {Chen}},\ }\href {\doibase http://dx.doi.org/10.1016/0375-9474(93)90197-6}
  {\bibfield  {journal} {\bibinfo  {journal} {Nuclear Physics A}\ }\textbf
  {\bibinfo {volume} {562}},\ \bibinfo {pages} {218 } (\bibinfo {year}
  {1993})}\BibitemShut {NoStop}%
\bibitem [{\citenamefont {Chen}(1997)}]{Chen97}%
  \BibitemOpen
  \bibfield  {author} {\bibinfo {author} {\bibfnamefont {J.-Q.}\ \bibnamefont
  {Chen}},\ }\href {\doibase http://dx.doi.org/10.1016/S0375-9474(97)00502-2}
  {\bibfield  {journal} {\bibinfo  {journal} {Nuclear Physics A}\ }\textbf
  {\bibinfo {volume} {626}},\ \bibinfo {pages} {686 } (\bibinfo {year}
  {1997})}\BibitemShut {NoStop}%
\bibitem [{\citenamefont {Zhao}\ and\ \citenamefont {Arima}(2014)}]{Zhao14}%
  \BibitemOpen
  \bibfield  {author} {\bibinfo {author} {\bibfnamefont {Y.}~\bibnamefont
  {Zhao}}\ and\ \bibinfo {author} {\bibfnamefont {A.}~\bibnamefont {Arima}},\
  }\href {\doibase http://doi.org/10.1016/j.physrep.2014.07.002} {\bibfield
  {journal} {\bibinfo  {journal} {Physics Reports}\ }\textbf {\bibinfo {volume}
  {545}},\ \bibinfo {pages} {1 } (\bibinfo {year} {2014})},\ \bibinfo {note}
  {nucleon-pair approximation to the nuclear shell model}\BibitemShut {NoStop}%
\bibitem [{\citenamefont {Fu}\ \emph {et~al.}(2013)\citenamefont {Fu},
  \citenamefont {Lei}, \citenamefont {Zhao}, \citenamefont {Pittel},\ and\
  \citenamefont {Arima}}]{Fu13}%
  \BibitemOpen
  \bibfield  {author} {\bibinfo {author} {\bibfnamefont {G.~J.}\ \bibnamefont
  {Fu}}, \bibinfo {author} {\bibfnamefont {Y.}~\bibnamefont {Lei}}, \bibinfo
  {author} {\bibfnamefont {Y.~M.}\ \bibnamefont {Zhao}}, \bibinfo {author}
  {\bibfnamefont {S.}~\bibnamefont {Pittel}}, \ and\ \bibinfo {author}
  {\bibfnamefont {A.}~\bibnamefont {Arima}},\ }\href {\doibase
  10.1103/PhysRevC.87.044310} {\bibfield  {journal} {\bibinfo  {journal} {Phys.
  Rev. C}\ }\textbf {\bibinfo {volume} {87}},\ \bibinfo {pages} {044310}
  (\bibinfo {year} {2013})}\BibitemShut {NoStop}%
\bibitem [{\citenamefont {Suzuki}\ and\ \citenamefont {Lee}(1980)}]{suzuki80}%
  \BibitemOpen
  \bibfield  {author} {\bibinfo {author} {\bibfnamefont {K.}~\bibnamefont
  {Suzuki}}\ and\ \bibinfo {author} {\bibfnamefont {S.~Y.}\ \bibnamefont
  {Lee}},\ }\href {\doibase 10.1143/PTP.64.2091} {\bibfield  {journal}
  {\bibinfo  {journal} {Prog.\ Theor.\ Phys.}\ }\textbf {\bibinfo {volume}
  {64}},\ \bibinfo {pages} {2091} (\bibinfo {year} {1980})}\BibitemShut
  {NoStop}%
\bibitem [{\citenamefont {Navr\'atil}\ \emph {et~al.}(1993)\citenamefont
  {Navr\'atil}, \citenamefont {Geyer},\ and\ \citenamefont {Kuo}}]{Navratil93}%
  \BibitemOpen
  \bibfield  {author} {\bibinfo {author} {\bibfnamefont {P.}~\bibnamefont
  {Navr\'atil}}, \bibinfo {author} {\bibfnamefont {H.}~\bibnamefont {Geyer}}, \
  and\ \bibinfo {author} {\bibfnamefont {T.}~\bibnamefont {Kuo}},\ }\href
  {\doibase http://dx.doi.org/10.1016/0370-2693(93)90148-B} {\bibfield
  {journal} {\bibinfo  {journal} {Physics Letters B}\ }\textbf {\bibinfo
  {volume} {315}},\ \bibinfo {pages} {1 } (\bibinfo {year} {1993})}\BibitemShut
  {NoStop}%
\bibitem [{\citenamefont {Van~Isacker}(2013)}]{Isacker13}%
  \BibitemOpen
  \bibfield  {author} {\bibinfo {author} {\bibfnamefont {P.}~\bibnamefont
  {Van~Isacker}},\ }\href {\doibase 10.1142/S0218301313300282} {\bibfield
  {journal} {\bibinfo  {journal} {International Journal of Modern Physics E}\
  }\textbf {\bibinfo {volume} {22}},\ \bibinfo {pages} {1330028} (\bibinfo
  {year} {2013})}\BibitemShut {NoStop}%
\bibitem [{\citenamefont {Otsuka}\ \emph {et~al.}(1978)\citenamefont {Otsuka},
  \citenamefont {Arima},\ and\ \citenamefont {Iachello}}]{Otsuka78}%
  \BibitemOpen
  \bibfield  {author} {\bibinfo {author} {\bibfnamefont {T.}~\bibnamefont
  {Otsuka}}, \bibinfo {author} {\bibfnamefont {A.}~\bibnamefont {Arima}}, \
  and\ \bibinfo {author} {\bibfnamefont {F.}~\bibnamefont {Iachello}},\ }\href
  {\doibase 10.1016/0375-9474(78)90532-8} {\bibfield  {journal} {\bibinfo
  {journal} {Nucl. Phys. A}\ }\textbf {\bibinfo {volume} {309}},\ \bibinfo
  {pages} {1} (\bibinfo {year} {1978})}\BibitemShut {NoStop}%
\bibitem [{\citenamefont {Skouras}\ \emph {et~al.}(1990)\citenamefont
  {Skouras}, \citenamefont {Van~Isacker},\ and\ \citenamefont
  {Nagarajan}}]{Skouras90}%
  \BibitemOpen
  \bibfield  {author} {\bibinfo {author} {\bibfnamefont {L.~D.}\ \bibnamefont
  {Skouras}}, \bibinfo {author} {\bibfnamefont {P.}~\bibnamefont
  {Van~Isacker}}, \ and\ \bibinfo {author} {\bibfnamefont {M.~A.}\ \bibnamefont
  {Nagarajan}},\ }\href {\doibase
  http://dx.doi.org/10.1016/0375-9474(90)90308-9} {\bibfield  {journal}
  {\bibinfo  {journal} {Nuclear Physics A}\ }\textbf {\bibinfo {volume}
  {516}},\ \bibinfo {pages} {255 } (\bibinfo {year} {1990})}\BibitemShut
  {NoStop}%
\bibitem [{\citenamefont {L\"owdin}(1950)}]{Lowdin50}%
  \BibitemOpen
  \bibfield  {author} {\bibinfo {author} {\bibfnamefont {P.}~\bibnamefont
  {L\"owdin}},\ }\href {\doibase 10.1063/1.1747632} {\bibfield  {journal}
  {\bibinfo  {journal} {The Journal of Chemical Physics}\ }\textbf {\bibinfo
  {volume} {18}},\ \bibinfo {pages} {365} (\bibinfo {year} {1950})},\ \Eprint
  {http://arxiv.org/abs/http://dx.doi.org/10.1063/1.1747632}
  {http://dx.doi.org/10.1063/1.1747632} \BibitemShut {NoStop}%
\bibitem [{\citenamefont {Carlson}\ and\ \citenamefont
  {Keller}(1957)}]{Carlson57}%
  \BibitemOpen
  \bibfield  {author} {\bibinfo {author} {\bibfnamefont {B.~C.}\ \bibnamefont
  {Carlson}}\ and\ \bibinfo {author} {\bibfnamefont {J.~M.}\ \bibnamefont
  {Keller}},\ }\href {\doibase 10.1103/PhysRev.105.102} {\bibfield  {journal}
  {\bibinfo  {journal} {Phys. Rev.}\ }\textbf {\bibinfo {volume} {105}},\
  \bibinfo {pages} {102} (\bibinfo {year} {1957})}\BibitemShut {NoStop}%
\bibitem [{\citenamefont {Mayer}(2002)}]{Mayer02}%
  \BibitemOpen
  \bibfield  {author} {\bibinfo {author} {\bibfnamefont {I.}~\bibnamefont
  {Mayer}},\ }\href {\doibase 10.1002/qua.981} {\bibfield  {journal} {\bibinfo
  {journal} {International Journal of Quantum Chemistry}\ }\textbf {\bibinfo
  {volume} {90}},\ \bibinfo {pages} {63} (\bibinfo {year} {2002})}\BibitemShut
  {NoStop}%
\bibitem [{\citenamefont {Poves}\ \emph {et~al.}(2001)\citenamefont {Poves},
  \citenamefont {S{\'a}nchez-Solano}, \citenamefont {Caurier},\ and\
  \citenamefont {Nowacki}}]{Poves01}%
  \BibitemOpen
  \bibfield  {author} {\bibinfo {author} {\bibfnamefont {A.}~\bibnamefont
  {Poves}}, \bibinfo {author} {\bibfnamefont {J.}~\bibnamefont
  {S{\'a}nchez-Solano}}, \bibinfo {author} {\bibfnamefont {E.}~\bibnamefont
  {Caurier}}, \ and\ \bibinfo {author} {\bibfnamefont {F.}~\bibnamefont
  {Nowacki}},\ }\href {\doibase 10.1016/S0375-9474(01)00967-8} {\bibfield
  {journal} {\bibinfo  {journal} {Nucl. Phys. A}\ }\textbf {\bibinfo {volume}
  {694}},\ \bibinfo {pages} {157} (\bibinfo {year} {2001})}\BibitemShut
  {NoStop}%
\bibitem [{\citenamefont {Caurier}\ \emph {et~al.}(2005)\citenamefont
  {Caurier}, \citenamefont {Mart{\'i}nez-Pinedo}, \citenamefont {Nowacki},
  \citenamefont {Poves},\ and\ \citenamefont {Zuker}}]{cau05}%
  \BibitemOpen
  \bibfield  {author} {\bibinfo {author} {\bibfnamefont {E.}~\bibnamefont
  {Caurier}}, \bibinfo {author} {\bibfnamefont {G.}~\bibnamefont
  {Mart{\'i}nez-Pinedo}}, \bibinfo {author} {\bibfnamefont {F.}~\bibnamefont
  {Nowacki}}, \bibinfo {author} {\bibfnamefont {A.}~\bibnamefont {Poves}}, \
  and\ \bibinfo {author} {\bibfnamefont {A.~P.}\ \bibnamefont {Zuker}},\ }\href
  {\doibase 10.1103/RevModPhys.77.427} {\bibfield  {journal} {\bibinfo
  {journal} {Rev. Mod. Phys.}\ }\textbf {\bibinfo {volume} {77}},\ \bibinfo
  {pages} {427} (\bibinfo {year} {2005})}\BibitemShut {NoStop}%
\bibitem [{\citenamefont {Dufour}\ and\ \citenamefont {Zuker}(1996)}]{duf96}%
  \BibitemOpen
  \bibfield  {author} {\bibinfo {author} {\bibfnamefont {M.}~\bibnamefont
  {Dufour}}\ and\ \bibinfo {author} {\bibfnamefont {A.~P.}\ \bibnamefont
  {Zuker}},\ }\href {\doibase 10.1103/PhysRevC.54.1641} {\bibfield  {journal}
  {\bibinfo  {journal} {Phys. Rev. C}\ }\textbf {\bibinfo {volume} {54}},\
  \bibinfo {pages} {1641} (\bibinfo {year} {1996})}\BibitemShut {NoStop}%
\bibitem [{\citenamefont {Elliott}\ and\ \citenamefont
  {White}(1980)}]{elliott80}%
  \BibitemOpen
  \bibfield  {author} {\bibinfo {author} {\bibfnamefont {J.~P.}\ \bibnamefont
  {Elliott}}\ and\ \bibinfo {author} {\bibfnamefont {A.~P.}\ \bibnamefont
  {White}},\ }\href@noop {} {\bibfield  {journal} {\bibinfo  {journal} {Phys.
  Lett. B}\ }\textbf {\bibinfo {volume} {97}},\ \bibinfo {pages} {169}
  (\bibinfo {year} {1980})}\BibitemShut {NoStop}%
\bibitem [{\citenamefont {Elliott}\ and\ \citenamefont
  {Evans}(1981)}]{elliott81}%
  \BibitemOpen
  \bibfield  {author} {\bibinfo {author} {\bibfnamefont {J.~P.}\ \bibnamefont
  {Elliott}}\ and\ \bibinfo {author} {\bibfnamefont {J.~A.}\ \bibnamefont
  {Evans}},\ }\href@noop {} {\bibfield  {journal} {\bibinfo  {journal} {Phys.
  Lett. B}\ }\textbf {\bibinfo {volume} {101}},\ \bibinfo {pages} {216}
  (\bibinfo {year} {1981})}\BibitemShut {NoStop}%
\bibitem [{\citenamefont {Thompson}\ \emph {et~al.}(1987)\citenamefont
  {Thompson}, \citenamefont {Elliott},\ and\ \citenamefont
  {Evans}}]{Thompson87}%
  \BibitemOpen
  \bibfield  {author} {\bibinfo {author} {\bibfnamefont {M.~J.}\ \bibnamefont
  {Thompson}}, \bibinfo {author} {\bibfnamefont {J.~P.}\ \bibnamefont
  {Elliott}}, \ and\ \bibinfo {author} {\bibfnamefont {J.~A.}\ \bibnamefont
  {Evans}},\ }\href {\doibase 10.1016/0370-2693(87)91563-2} {\bibfield
  {journal} {\bibinfo  {journal} {Phys. Lett.}\ }\textbf {\bibinfo {volume}
  {B195}},\ \bibinfo {pages} {511} (\bibinfo {year} {1987})}\BibitemShut
  {NoStop}%
\bibitem [{\citenamefont {Afanasjev}(2012)}]{Afanasjev12}%
  \BibitemOpen
  \bibfield  {author} {\bibinfo {author} {\bibfnamefont {A.~V.}\ \bibnamefont
  {Afanasjev}},\ }\href@noop {} {\  (\bibinfo {year} {2012})},\ \Eprint
  {http://arxiv.org/abs/1205.2134} {arXiv:1205.2134 [nucl-th]} \BibitemShut
  {NoStop}%
\bibitem [{\citenamefont {Evans}\ \emph {et~al.}(1993)\citenamefont {Evans},
  \citenamefont {Long},\ and\ \citenamefont {Elliott}}]{Evans93}%
  \BibitemOpen
  \bibfield  {author} {\bibinfo {author} {\bibfnamefont {J.~A.}\ \bibnamefont
  {Evans}}, \bibinfo {author} {\bibfnamefont {G.}~\bibnamefont {Long}}, \ and\
  \bibinfo {author} {\bibfnamefont {J.~P.}\ \bibnamefont {Elliott}},\ }\href
  {\doibase http://dx.doi.org/10.1016/0375-9474(93)90151-M} {\bibfield
  {journal} {\bibinfo  {journal} {Nuclear Physics A}\ }\textbf {\bibinfo
  {volume} {561}},\ \bibinfo {pages} {201 } (\bibinfo {year}
  {1993})}\BibitemShut {NoStop}%
\bibitem [{\citenamefont {Evans}\ \emph {et~al.}(1995)\citenamefont {Evans},
  \citenamefont {Elliott}, \citenamefont {Lac},\ and\ \citenamefont
  {Long}}]{Evans95}%
  \BibitemOpen
  \bibfield  {author} {\bibinfo {author} {\bibfnamefont {J.~A.}\ \bibnamefont
  {Evans}}, \bibinfo {author} {\bibfnamefont {J.~P.}\ \bibnamefont {Elliott}},
  \bibinfo {author} {\bibfnamefont {V.~S.}\ \bibnamefont {Lac}}, \ and\
  \bibinfo {author} {\bibfnamefont {G.~L.}\ \bibnamefont {Long}},\ }\href
  {\doibase http://dx.doi.org/10.1016/0375-9474(95)00326-V} {\bibfield
  {journal} {\bibinfo  {journal} {Nuclear Physics A}\ }\textbf {\bibinfo
  {volume} {593}},\ \bibinfo {pages} {85 } (\bibinfo {year}
  {1995})}\BibitemShut {NoStop}%
\bibitem [{\citenamefont {Barea}\ \emph
  {et~al.}(2015{\natexlab{b}})\citenamefont {Barea}, \citenamefont {Kotila},\
  and\ \citenamefont {Iachello}}]{bar15a}%
  \BibitemOpen
  \bibfield  {author} {\bibinfo {author} {\bibfnamefont {J.}~\bibnamefont
  {Barea}}, \bibinfo {author} {\bibfnamefont {J.}~\bibnamefont {Kotila}}, \
  and\ \bibinfo {author} {\bibfnamefont {F.}~\bibnamefont {Iachello}},\ }\href
  {\doibase 10.1103/PhysRevD.92.093001} {\bibfield  {journal} {\bibinfo
  {journal} {Phys. Rev. D}\ }\textbf {\bibinfo {volume} {92}},\ \bibinfo
  {pages} {093001} (\bibinfo {year} {2015}{\natexlab{b}})}\BibitemShut
  {NoStop}%
\bibitem [{\citenamefont {Talmi}(1993)}]{talmi93}%
  \BibitemOpen
  \bibfield  {author} {\bibinfo {author} {\bibfnamefont {I.}~\bibnamefont
  {Talmi}},\ }\href@noop {} {\emph {\bibinfo {title} {Simple Models of Complex
  Nuclei : The Shell Model and Interacting Boson Model}}}\ (\bibinfo
  {publisher} {Harwood Academic Publishers},\ \bibinfo {address} {Chur,
  Switzerland; Langhorne, Pa., USA},\ \bibinfo {year} {1993})\BibitemShut
  {NoStop}%
\end{thebibliography}

%

\end{document}